\RequirePackage{fix-cm}
\RequirePackage{rotating}

\documentclass[smallextended]{svjour3}       
\smartqed  

\usepackage{rotating}

\usepackage{csquotes}
\usepackage{booktabs} 
\usepackage{framed}
\usepackage{graphicx}
\usepackage{amssymb}
\usepackage{tabularx}
\usepackage{ltablex}
\usepackage{graphicx}
\usepackage{url}
\usepackage{tabulary}
\usepackage{multirow}
\usepackage{enumitem}
\usepackage{hyperref}
\usepackage{hypcap}
\usepackage{lscape}
\usepackage[normalem]{ulem}
\useunder{\uline}{\ul}{}
\usepackage{longtable}
\usepackage{color}
\usepackage{hyperref}

\usepackage{booktabs}
\usepackage[flushleft]{threeparttable}
\usepackage{microtype}

\usepackage{placeins}

\usepackage{siunitx}  
\usepackage{textgreek}

\usepackage{flushend}

\usepackage{lscape}

\usepackage{rotating}

\begin{document}

\title{Do Agile Scaling Approaches Make A Difference? \\An Empirical Comparison of Team Effectiveness Across Popular Scaling Approaches}

\titlerunning{Do Agile Scaling Approaches Make A Difference?}

\author{Christiaan Verwijs      \&
        Daniel Russo
}


\institute{C. Verwijs \at The Liberators BV, The Netherlands
\and
D. Russo  \at
            Corresponding author.\\
              Aalborg University, Department of Computer Science, Copenhagen, Denmark \\
              Tel.: (+45) 9940 7765\\
              \email{daniel.russo@cs.aau.dk}           
}

\date{Received: DD Month YEAR / Accepted: DD Month YEAR}
\maketitle

\begin{abstract}
In the context of the prevalent use of Agile methodologies, organizations are grappling with the challenge of scaling development across numerous teams. This has led to the emergence of diverse scaling strategies, from complex ones such as ``SAFe", to more simplified methods e.g., ``LeSS", with some organizations devising their unique approaches. While there have been multiple studies exploring the organizational challenges associated with different scaling approaches, so far, no one has compared these strategies based on empirical data derived from a uniform measure. This makes it hard to draw robust conclusions about how different scaling approaches affect Agile team effectiveness. Thus, the objective of this study is to assess the effectiveness of Agile teams across various scaling approaches, including ``SAFe", ``LeSS", ``Scrum of Scrums", and custom methods, as well as those not using scaling. The assessment focuses initially on responsiveness, stakeholder concern, continuous improvement, team autonomy, management approach, and overall team effectiveness, followed by an evaluation based on stakeholder satisfaction regarding value, responsiveness, and release frequency. To achieve this, we performed a comprehensive survey involving 15,078 members of 4,013 Agile teams to measure their effectiveness, combined with satisfaction surveys from 1,841 stakeholders of 529 of those teams. We conducted a series of inferential statistical analyses, including Analysis of Variance and multiple linear regression, to identify any significant differences, while accounting for team experience and organizational size. The findings of the study revealed some significant differences, but their magnitude and effect size were considered too negligible to have practical significance. In conclusion, the choice of Agile scaling strategy does not markedly influence team effectiveness, and organizations are advised to choose a method that best aligns with their previous experiences with Agile, organizational culture, and management style.

\end{abstract}

\section{Introduction}
\label{sec:Introduction}

Agile methodologies have revolutionized software development by championing iterative processes, adaptability, and a stakeholder-focused approach, ensuring the efficient delivery of high-quality products. This transformation is evident as 80\% of organizations now predominantly employ Agile processes in their operations~\cite{version1stateofagile}. While in its initial adoption stage, Agile was largely confined to single teams~\cite{Strode2012}, there was a notable absence of guidance on how to scale Agile practices across multiple teams. However, the success of Agile at the team level has not only expanded its application beyond the realm of software development but has also pushed its implementation in large-scale settings~\cite{dingsoyr2018}.

In recent years, Agile Scaling frameworks have become increasingly popular to address this gap~\cite{mishra2011}, including the Scaled Agile Framework (``SAFe'')~\cite{SAFewebsite}, Large-Scale Scrum (``LeSS'')~\cite{LESSWebsite}, ``Nexus''~\cite{NexusWebsite}, Scrum@Scale and ``Scrum of Scrums''~\cite{sutherland2001inventing,schwaber2004agile}. ``SAFe'' has become the most popular with 53\% of organizations opting for it~\cite{version1stateofagile}, followed by 28\% Scrum@Scale (often referred to as ``Scrum of Scrums''). ``LeSS'' is adopted by 6\% of organizations and Nexus by 3\%. ``SAFe'' has also been identified as the most popular in scholarly investigations~\cite{alqudah2016,SAFe2018litrev,Conboy19}. However, many organizations also develop their own approaches to scale Agile development across many teams~\cite{Edison22}. These results show that organizations pick different solutions to scale, without a universally agreed best practice for software teams.

Among the myriad approaches, ``SAFe'' is often viewed as the most complex~\cite{Ebert17}. Some anecdotal evidence suggests it is not well-received within the professional Agile community. For example, a survey conducted in the industrial sector with 505 professionals using Agile methods found that a notable number of participants were unlikely to recommend the methods they were using~\cite{wolpers2023safe}. Furthermore, we report a dedicated website that collects criticisms of ``SAFe'' from Agile experts as well as case studies~\cite{safedelusion}.

Nevertheless, the interest in Agile scaling approaches has sparked an interest of practitioners and researchers alike. In particular, 136 articles have been published in 46 venues by more than 200 authors between 2009 and 2019, accessible through IEEE Xplore, ACM Digital Library, Science Direct, Web of Science, and AIS eLibrary~\cite{ULUDAG22}. The aim of this study is to investigate how scaling approaches like ``SAFe'', ``LeSS'', ``Scrum of Scrums'', as well as custom approaches, impact the effectiveness of Agile teams and the satisfaction of their stakeholders. Henceforth, we frame our research question (RQ) as follows:\\

RQ: \textit{To what extent are the effectiveness of Agile teams and the satisfaction of their stakeholders influenced by the Agile scaling approach in use?}\\

To answer our research question, we performed a cross-sectional study with 15,078 team members aggregated into 4,013 Agile teams. We compared their overall effectiveness and the quality of core processes of Agile teams as operationalized by Verwijs \& Russo~\cite{verwijs2023theory}. Furthermore, we analyzed the evaluations of 1,746 stakeholders (e.g., users, customers, and internal stakeholders) for 544 of those teams. Analysis of Variance (ANOVA) and linear regression were used to identify significant differences between scaling approaches, with and without controlling for the experience of teams with Agile and the size of organizations.

Our research revealed that small, but statistically significant differences among scaling approaches. However, their effect sizes were too small to be practically relevant. In essence, the choice of scaling approach seems to have a negligible impact on team effectiveness and stakeholder satisfaction. Notably, among the control variables, a team's experience with Agile emerged as a more influential factor.

In the remainder of this paper, we describe the related work in Section~~\ref{sec:related}.
We then discuss our research design in Section~~\ref{sec:researchdesign} and report the results of our analyses in Section~~\ref{sec:analysis}. Finally, we discuss the implications for research and practice along with the study limitations in Section~~\ref{sec:Discussion} and draw our conclusion by outlying future research directions in Section~~\ref{sec:Conclusion}.

\section{Related Work}
\label{sec:related}
Organizations often engage in multi-year software engineering projects that involve the coordination of work done by many teams, either regionally, globally, or both~\cite{Ebert17}. With the rise of Agile methods and their collaborative, iterative, and human-oriented approach to software engineering, organizations are increasingly seeking ways to apply Agile principles at scale. Agile methods like Scrum and XP initially focused on intra-team collaboration and offered little guidance on how to apply it across many teams~\cite{AgileManifesto2001,schwaber2020scrum}. While this works well in small organizations or efforts that involve few teams, many challenges have been identified in the application to large-scale efforts~\cite{maples2009}. 

Empirical research on how companies can do large-scale transformations and processes has been scarce~\cite{Ebert17}, with some exceptions~\cite{paasivaara2018large,russo2021agile}. Several researchers have tried to define when a project is considered large-scale, and a taxonomy of the scale of Agile has been developed. Dingsøyr et al.~\cite{dingsoyr2014} state that the cost of a project is not a sufficient criterion for large-scale, as some projects might involve hardware procurement, which differs in price related to the specific country. The reliable factor in defining large-scale projects is the number of teams the practitioners are divided into, where 2-9 teams are considered large-scale, and over ten teams are considered very large-scale~\cite{dingsoyr2014}.

\subsection{Scaled Agile Framework (``SAFe'')}
The most popular Agile scaling approach to date is the Scaled Agile Framework (``SAFe'') with an approximate market share of 53\%, according to a recent industry survey~\cite{version1stateofagile}. It aims to enable large-scale software and product development by applying Agile principles at the enterprise level~\cite{SAFewebsite}. Developed by Dean Leffingwell, the framework combines elements of Agile, Scrum, Lean, and related methodologies and is organized into three levels - Team, Program, and Portfolio. ``SAFe'' emphasizes continuous improvement, collaboration, and alignment between teams and offers a range of tools and practices, including Agile Release Trains (ARTs), PI planning, and DevOps, to support these objectives. The framework is designed to help organizations achieve faster time-to-market, higher quality, and greater efficiency in their software and systems development efforts~\cite{alqudah2016,SAFewebsite}. However, some practitioners perceive ``SAFe'' as complex due to its attempt to incorporate all best practices and its failure to explain how to scale down~\cite{safedelusion}. SAFe includes many role definitions, which make managers feel comfortable, and uses Scrum practices at the team level, with the opportunity to use Kanban, while applying specific roles such as product manager, system architect, and deployment team~\cite{Ebert17,SAFewebsite}

Putta, Paasivaara \& Lassenius investigated the challenges and benefits of ``SAFe'' with a multivocal literature study. They reviewed six academic studies and 47 non-peer-reviewed case studies provided by the developers of ``SAFe'' to identify patterns in benefits and challenges. The most common business benefits of adopting ``SAFe'' were transparency, alignment, quality, time to market, predictability, and productivity~\cite{Putta18}. The authors note that these benefits were named specifically only in the case studies, but not in the academic studies. This difference is likely because the developers of ``SAFe'' focused on the business benefits in the writing of the case studies. The most common challenges mentioned in the reviewed studies were; resistance to change, moving away from Agile, first PI planning, controversies with the framework, Agile Release Train challenges, staffing roles, and GSD (Guided Self-Determination) challenges~\cite{Putta18}.

Ciancarini et al.~\cite{Ciancarini22} conducted a multivocal literature review that focused on the challenges in the adoption of ``SAFe''. The study covers three main research areas; to identify the success factors, to uncover implementation issues, and to discover the effects. The success factors for ``SAFe'' include Leadership Support in transformation, Communication between Layers, Support from Teammates, and Trust between Teams. Therefore it is beneficial for top management to both support and understand ``SAFe'', along with a shared commitment to adopt ``SAFe'' in all parts of the organization for the framework to succeed. Ciancarini et al. also performed interviews with 25 respondents from 17 organizations to gain a deeper understanding of the challenges. They found that practitioners initially experience ``SAFe'' as very complex and overwhelming. However, the approach becomes more effective after the initial stage. According to the study, the most significant reported benefits of ``SAFe'' relate to better company management, such as increased productivity, shared vision, and coordination of work. The most commonly identified challenges are that ``SAFe'' requires a major commitment on all levels of the organization, resources in the form of time, and that ``SAFe'' may inhibit Agility when improperly practiced and misunderstood by management~\cite{Ciancarini22}.

\subsection{Large-Scale Scrum (``LeSS'')}
Large-Scale Scrum (``LeSS'') was developed by Craig Larman \& Bas Vodde~\cite{LESSWebsite}. It aims to scale Scrum, lean, and Agile development principles to large product groups. ``LeSS'' remains conceptually close to the Scrum Framework and is more lightweight than ``SAFe''~\cite{kalenda2018scaling}. In ``LeSS'', all Scrum Teams start and end their Sprints at the same time and deliver one potentially shippable increment together in that time. Each Sprint begins with a shared Sprint Planning and ends with a shared Sprint Review and Sprint Retrospective. Work is pulled from a shared Product Backlog and is managed by a single Product Owner. While ``LeSS'' is recommended for up to 8 Scrum teams, multiple ``LeSS'' frameworks can be stacked to accommodate larger numbers in ``LeSS Huge''. 

\subsection{Scrum of Scrums and Scrum@Scale}
``Scrum of Scrums'' is one of the earliest approaches to scale Agile development across multiple teams~\cite{sutherland2001inventing,schwaber2004agile}. It is more aptly described as a practice that is applied on top of the Scrum framework than a full framework in its own right~\cite{kalenda2018scaling}. It builds on the ``Daily Scrum'' that is held every 24 hours by each Scrum team to coordinate work between its members and is timeboxed to 15 minutes. Each team then sends one member to a ``Daily Scrum'' that is held every 24 hours to coordinate work across teams and manage dependencies~\cite{schwaber2004agile}. Although the ``Scrum of Scrums'' is recommended for settings with up to 10 teams, multiple levels of ``Scrum of Scrums'' can accommodate larger scales~\cite{sutherland2001inventing}. The ``Scrum of Scrums'' is the core practice of the Scrum@Scale-framework~\cite{ScrumAtScaleWebsite}, although it adds Scaled Retrospectives, a Scrum Master for the facilitation of the ``Scrum of Scrums'' and a Scrum team to remove impediments called ``Executive Action Team''. Like ``LeSS'', ``Scrum@Scale'' is more lightweight than ``SAFe''.

\subsection{Other approaches}
More approaches have been developed to scale Agile development across multiple teams. We describe a selection below with the aim of illustrating the broadness of the landscape.

``Disciplined Agile'' (DA) was developed by Scott Ambler \& Mark Lines~\cite{ambler2020introduction}. It provides a more comprehensive multi-phased process model for scaled Agile delivery that also includes expert roles for technical architecture, testing, domain expertise, and integration. ``Nexus'' is another lightweight scaling approach developed by Ken Schwaber~\cite{bittner2017nexus} that remains conceptually close to the Scrum framework. It introduces scaled versions of the Sprint Planning, Sprint Review, and Sprint Retrospectives that are held by up to 9 teams. A ``Nexus Integration Team'' is introduced to coordinate the integration of work between teams and provide training, support, and coaching. Another perspective on scaling was provided by Henrik Kniberg in the ``Spotify Model''~\cite{kniberg2014spotify}. It is not a framework but rather describes how Spotify organized the scaling of its development and its culture across many teams in the early 2010s. The last specific approach we will discuss here is ``Recipe for Agile Governance'' (RAGE) by Kevin Thompson~\cite{thompson2013recipes}. It provides a set of practices and roles drawn from Scrum and Lean to provide guidance at the project-, program- and portfolio levels in large enterprises. 

Finally, Conboy \& Carroll~\cite{Conboy19} observe that predominant corporations such as Dell, Accenture, and Intel frequently formulate their unique scaling strategies. This customization is aimed at ensuring a more harmonious integration with the prevailing organizational culture and structures and to more effectively comply with regulatory mandates~\cite{kostic2017challenges}. This trend is not exclusive to these entities; mission-critical organizations, which are often subject to stringent security prerequisites, also exhibit a preference for tailored development to meet their specific needs~\cite{messina2016new,ciancarini2018agile,russo2018contracting}.

\subsection{Reviews of Agile scaling approaches}
Almeida \& Espinheira~\cite{Almeida21} studied the performance of six large-scale Agile frameworks on 15 assessment criteria, including the level of control, customer involvement, and technical complexity. Their review included ``Disciplined Agile'', ``LeSS'', ``Nexus'', ``SAFe'', ``Scrum@Scale'', and Spotify’s Agile Scaling Model. None performed better on all dimensions. The authors argue that the optimal approach for organizations is to adopt the framework most similar to their current mindset~\cite{Almeida21}.

A systematic literature review by Edison, Wang \& Conboy~\cite{Edison22} also identified challenges common to ``SAFe'', ``Scrum@Scale'', ``Disciplined Agile'', ``Spotify’s Agile Scaling Model'', and ``LeSS''. They collected 191 studies across 134 different organizations that considered one or more of these approaches in primary studies published between 2003 and 2019. The authors identified 31 challenges grouped into nine distinct areas when scaling Agile: inter-team coordination, customer collaboration, architecture, organizational structure, method adoption, change management, team design, and project management. Based on 191 studies they reviewed, they conclude that none of these challenges are unique to specific large-scale methodologies. According to these authors, opting for a custom approach may lead to slightly more challenges~\cite{Edison22}. While this study aimed to identify adoption patterns and not compare the different methodologies, the findings support those of Almeida \& Espinheira by underlining the importance of context when evaluating the effectiveness of Agile frameworks~\cite{Almeida21,Edison22}. A framework that performs optimally in one setting can perform ineffectively in another. 

The pattern that emerges from the literature is that one scaling approach is not clearly better than the others. Although there seems to be a preference for simpler approaches by practitioners, lightweight approaches like ``LeSS'', ``Scrum of Scrums'' or ``Nexus'' do not appear to be categorically better than more complex approaches like ``SAFe'' or ``Disciplined Agile''. Instead, contextual variables seem to be more decisive in determining what is best for an organization. However, the aim of the aforementioned studies was to identify challenges and success factors across primary studies (e.g., case studies) of scaling approaches as implemented in case organizations. The qualitative nature of such data does not allow statistical generalization nor does it provide a comparison on equal grounds. To date, no empirical study has been performed that directly compares scaling approaches quantitatively on key metrics~\cite{Ebert17}. This study attempts to address that gap. Such a study provides empirical support for the patterns identified in the aforementioned investigations. Moreover, it brings clarity to how various scaling approaches perform, highlights potential variables that influence that performance, and offers evidence-based recommendations to the ongoing debate among practitioners~\cite{wolpers2023safe,safedelusion}.

\section{Research Design}
\label{sec:researchdesign}
To address our research question, we conducted a comprehensive survey targeting both teams and their associated stakeholders. We employed Analysis of Variance (ANOVA)~\cite{hair2019multivariate} and multiple linear regression~\cite{hair2019multivariate} to compare the results between different scaling frameworks. This section discusses the research hypotheses Sec.~\ref{sec:hypotheses}, the sample (Sec.~\ref{sec:participants}), measurement instruments (Sec.~\ref{sec:measurements}), and method of analysis (Sec.~\ref{sec:analysis}).

\subsection{Research Hypotheses}
\label{sec:hypotheses}
This study contributes to existing research by being the first to use a quantitative approach to empirically compare the results on a consistent set of measures across different scaling approaches. We will do so through the lens of ``Team Effectiveness'' and the team-level processes that give rise to it. Hackman~\cite{hackman1976design} defines \textit{Team effectiveness} as \textit{``the degree to which a team meets the expectations of the quality of the outcome''}~\cite{hackman1976design}. 

Verwijs \& Russo~\cite{verwijs2023theory} developed and tested a five-factor model of team effectiveness with data from 1,978 Scrum teams. This model defines team effectiveness as two perspectives on the quality of the outcomes. The first perspective represents the satisfaction of stakeholders (e.g., clients, customers, and users) with the outcomes (stakeholder satisfaction). The second captures the satisfaction of team members with their team and its outcomes (team morale). The model also identifies five higher-order core processes at the team level that have been shown to determine the effectiveness of Scrum teams. 
The first factor is \textit{Responsiveness}. It reflects the ability of teams to respond quickly to emerging needs and requirements by stakeholders. Its lower-order processes are release frequency, release automation, and refinement. \textit{Stakeholder Concern} captures to which extent teams understand what is important to their stakeholders and work to clarify it. Its lower-order processes include stakeholder collaboration, shared goals, sprint review quality, and value focus. The third factor is \textit{Continuous Improvement} and captures the degree to which teams engage in a process of continuous improvement and feel the safety to do so. It is composed of the lower-order processes of psychological safety, concern for quality, shared learning, metric usage, and learning environment. \textit{Team autonomy} is the fourth factor and reflects the latitude of teams to manage their own work. It is composed of the lower-order processes of self-management and cross-functionality. The fifth process represents management support.

This study primarily investigates one scaling approach of higher complexity (``SAFe''), and two approaches of lower complexity (``LeSS'' and ``Scrum of Scrums''). A separate category is custom approaches to scaling that are developed by organizations internally. Consistent with the pattern from other comparative studies, we do not expect to find substantial differences between scaling approaches on team effectiveness and the five core processes that contribute to it. Thus, we hypothesize:\\

Hypothesis 1 (H1). \textit{Between scaling approaches, Agile teams are similar in terms of their responsiveness (H1a), concern for stakeholders (H1b), their ability to improve continuously (H1c), autonomy (H1d), management support (H1e), and their overall effectiveness (H1f).}\\

One limitation of the study by Verwijs \& Russo~\cite{verwijs2023theory} is that it measured stakeholder satisfaction indirectly through the perception of team members. Such measures are susceptible to a ``halo effect''~\cite{mathieu2008team} where teams that feel they are doing well may inflate their perceived satisfaction of stakeholders. To address this, we aim to directly measure the satisfaction of the stakeholders of teams with the responsiveness, release frequency, and quality of what is delivered by teams. Similarly to H1, we do not expect substantial differences in stakeholder satisfaction based on the scaling approach alone:\\

Hypothesis 2 (H2). \textit{Between scaling approaches, the satisfaction of stakeholders is similar for quality (H2a), responsiveness (H2b), and value (H2c).}\\

\subsubsection{Control Variables}
This study includes two control variables to account for alternative explanations. The first control variable concerns the experience that teams have with Agile. Since lightweight approaches prescribe less than more complex ones, it is reasonable to assume that teams that are less experienced with Agile may struggle more with lightweight frameworks whereas the reverse may be true for very experienced teams with highly prescriptive frameworks. Thus, we will control for the experience that teams have with Agile when comparing scaling approaches. 

The second control variable concerns the size of the organization a team is part of. Large organizations may be more inclined to opt for enterprise-oriented frameworks like ``SAFe'' or a custom approach, whereas smaller organizations may prefer the simplicity of ``LeSS'' or ``Scrum of Scrums''. Since the size of an organization itself may influence the effectiveness of teams, we will control for it in this study.

\subsection{Participants}
\label{sec:participants}
Data collection was performed between September 2021 and September 2023 through a public online survey\footnote{The GDPR-compliant survey has been designed so that teams can self-assess their Agile development process. It is available at the following URL: www.scrumteamsurvey.org.}. 15,078 members of 4,013 Agile teams participated in the survey, as well as 1,841 stakeholders of 529 of those teams. Due to the public nature of the survey, we were unable to calculate a response rate.

Public surveys are susceptible to response bias due to the self-selection of participants~\cite{meade2012identifying}. We employed several strategies outlined in the literature to reduce this threat to the validity of this study~\cite{meade2012identifying}. First, we ensured that team members and stakeholders could participate anonymously and emphasized this anonymity in our communication. Second, we encouraged honest answers by providing teams with a detailed team-level report and relevant feedback for their team upon completion. Third, to ensure a higher response rate from stakeholders, we provided teams with a mechanism to invite stakeholders themselves by sharing a link to a standardized questionnaire for stakeholders. Fourth, we removed all survey participants with a completion time below the 5\% percentile of the completion times for their segment (team members or stakeholders) as well as all participants that entered very few questions (less than ten for team members, and also less than ten for stakeholders). The composition of our sample is shown in table~\ref{tab:samplecomposition}.

\begin{table}[!ht]
\centering
\small
\caption{Composition of the sample}
\label{tab:samplecomposition}
\begin{tabularx}{0.75\textwidth}{@{}lll@{}}
\toprule
\textbf{Variable} & \textbf{Category} & \textbf{N (\%)} \\ \midrule
Respondents & Total & 15,078 \\
 & Team members & 12,534 \\
 & Stakeholders & 1,841 \\
Teams & Total & 4,013 \\
 & With stakeholder reports & 529 \\
Respondents per team & 1 respondent & 2,372 (59.1\%) \\
 & 2-4 respondents & 446 (11.1\%) \\
 & 5-8 respondents & 930 (23.2\%) \\
 & 9+ respondents & 265 (6.6\%) \\
Scaling Framework & Scale Agile Framework (SAFe) & 588 (14.7\%) \\
 & Large-Scale Scrum (LESS) & 142 (3.5\%) \\
 & Scrum of Scrums & 875 (21.8\%) \\
 & Homegrown Approach & 778 (19.4\%) \\
 & Other approach & 261 (6.5\%) \\
 & None & 1,369 (34.1\%) \\
Product Type & Product for internal users & 2,174 (54.2\%) \\
 & Product for external users / customers & 1,751 (43.6\%) \\
 & Unknown & 88 (2.2\%) \\
Scrum Team Size & 1-4 members & 174 (4.3\%) \\
 & 5-10 members & 2,807 (69.9\%) \\
 & 11-16 members & 777 (19.4\%) \\
 & \textgreater{}16 members & 173 (4.3\%) \\
 & Unknown & 82 (2\%) \\
Scrum Team Experience & Very low & 178 (4.4\%) \\
 & Low & 750 (18.7\%) \\
 & Moderate & 1,639 (40.8\%) \\
 & High & 1,446 (36\%) \\
Organization Sector & Technology And Telecommunications & 1,437 (35.8\%) \\
 & Financial & 665 (16.6\%) \\
 & Government & 208 (5.2\%) \\
 & Other & 1,703 (24.6\%) \\
Organization Size & 1-50 employees & 367 (9.1\%) \\
 & 51-500 employees & 1,324 (33\%) \\
 & 501-5.000 employees & 1,158 (28.9\%) \\
 & \textgreater{}5.000 employees & 947 (23.6\%) \\
 & Unknown & 217 (5.4\%) \\
Region & Europe & 1,651 (41.1\%) \\
 & North America & 396 (9.9\%) \\
 & Asia \& Oceania & 332 (8.3\%) \\
 & South America & 129 (3.2\%) \\
 & Africa \& Middle East & 112 (2.8\%) \\
 & Other \& Global & 1,393 (34.7\%) \\
Stakeholder Type & User (uses outcome of team) & 927 (50.4\%) \\
 & Customer (pays for outcome of teams) & 363 (19.7\%) \\
 & User and customer (both) & 307 (16.7\%) \\
 & Other & 193 (10.5\%) \\
 & Unknown & 51 (2.8\%) \\ \bottomrule
\end{tabularx}
\end{table}

\subsection{Measurements}
\label{sec:measurements}

\textbf{Scaling approach}: The scaling approach in use was measured at the team level with a single categorical item in the questionnaire (see also~\ref{sec:appendix:questionnaire}). This item was asked once of the Scrum Master, Product Owner, or manager who initiated the questionnaire for their team. The options included the scaling approaches ``SAFe'', ``LeSS'', ``Scrum of Scrums'', ``Custom approach'', one item to capture other scaling approaches (``Other approach''), and one option to indicate no scaling. We choose not to provide an exhaustive list of all potential scaling approaches (i.e. ``Nexus'', ``Disciplined Agile'', ``Spotify Model'' and ``RAGE'') and focus on those with larger market share~\cite{version1stateofagile} due to concerns that an exhaustive list might overwhelm participants. 

\textbf{Team effectiveness}: Team effectiveness was operationalized through a composite scale developed by Verwijs \& Russo~\cite{verwijs2023theory}. This scale measures two perspectives on the quality of the outcomes with two sub-scales. The first sub-scale measures the satisfaction of stakeholders (e.g., clients, customers, and users) with the outcomes (stakeholder satisfaction). The second sub-scale measures the satisfaction of team members with their team and its outcomes (team morale). Reliability analysis showed that the composite scale ($\alpha=.881$) was consistently measured across participants. The score for team effectiveness represents the mean-based average of the score for both sub-scales for each participating team member.

Furthermore, we measured five core processes that have been shown to predict the effectiveness of Agile teams by Verwijs \& Russo~\cite{verwijs2023theory}. \textit{Responsiveness} was operationalized with 8 Likert questions (1-7) ($\alpha=.847$) that represented the three sub-scales release frequency, release automation, and refinement. \textit{Stakeholder Concern} was operationalized with 13 Likert questions (1-7) ($\alpha=.914$) that represented four sub-scales: stakeholder collaboration, shared goals, sprint review quality, and value focus. The third factor is \textit{Continuous Improvement} was operationalized with 19 questions ($\alpha=.926$) from five sub-scales: psychological safety, concern for quality, shared learning, metric usage, and learning environment. \textit{Team autonomy} was measured with 5 questions ($\alpha=.819$) Likert questions (1-7) and consisted of two subs-scales; self-management and cross-functionality. Finally, management support is the final core indicator and it was measured with 2 questions ($\alpha=.848$). A heterotrait-monotrait (HTMT) analysis and Confirmatory Factor Analysis (CFA) were performed in~\cite{verwijs2023theory} and showed good discriminant and convergent validity.

Table~\ref{tab:scalesteameffectiveness} summarizes the scales, the number of items used for their operationalization, and their reliability (Cronbach's Alpha). The full questionnaire is provided in~\ref{sec:appendix:questionnaire}.

\begin{table*}[!ht]
\small
\centering
\caption{Scales used in the survey study to measure team effectiveness and five core indicators, number of items, and reliability (Cronbach's Alpha) based on respondent-level response data ($N=15,078$)}
\label{tab:scalesteameffectiveness}
\begin{tabularx}{\textwidth}{@{}lXll@{}}
\toprule
\textbf{Construct variable} & \textbf{\# Items} & \textbf{Alpha} \\ 
\midrule
Responsiveness & 8 & .847 \\
Stakeholder Concern & 13 & .914 \\
Continuous improvement & 19 & .926 \\
Team Autonomy & 5 & .819 \\
Management Support & 2 & .848 \\
Team Effectiveness & 6 & .881 \\ 
\bottomrule
\end{tabularx}
\end{table*}

\textbf{Stakeholder satisfaction}: The satisfaction reported by stakeholders was operationalized with a multidimensional construct consisting of four sub-scales of Likert questions (1-7). All scales were created by the authors. The data collection platform used for this study allowed teams and their Product Owners to invite relevant stakeholders themselves. Consequently, stakeholder evaluations were provided by 1,841 stakeholders for 529 teams.

The first sub-scale measured the \textit{satisfaction with value} and consisted of five items, including ``I am satisfied with the value that this team delivers'', ``What this team delivers is of high quality'' and ``When the team delivers a new version, it is usually free of serious bugs''. \textit{Satisfaction with responsiveness} operationalized how satisfied stakeholders are with the responsiveness of a team. It contained four questions, including ``When I have an idea or suggestion, members of the team are available to listen to me.'' and ``I frequently meet or interact with members of this team''. The third subs-scale measured the \textit{satisfaction with release frequency} with three items, including ``This team frequently delivers new versions'' and ``I am satisfied with how often new versions are released''.

A confirmatory factor analysis showed that all items loaded on their expected factors. Oblimin rotation was used to allow for correlations between related components. The 3 extracted components explained 57.3\% of the data variability. Individual item-factor loadings are reported 
in~\ref{sec:appendix:regressionanalysis}.

We assessed the discriminant validity of our three scales for stakeholder satisfaction with a heterotrait-monotrait (HTMT) analysis with a plugin for AMOS~\cite{gaskin2016master} and following the recommended process~\cite{hair2019multivariate,henseler2015new}. This ratio between trait correlations and within trait correlations should remain below $R=.90$ to indicate good discriminant validity from other constructs in different settings. This was the case for all scales.
We assessed convergent validity by inspecting composite reliability (CR) and average extracted variance (AVE). The AVE remained above the rule of thumb of $>.50\%$~\cite{hair2019multivariate} for all pairs of factors, ranging between .564 and .783. The CR was equal to or above the threshold of .7~\cite {hair2019multivariate} for all scales.

The resulting reliability (Cronbach's Alpha) of the three scales was satisfactory ($\textbf{Alpha} > .7$) and is reported in Table~\ref{tab:scalesstakeholdersatisfaction}. The full questionnaire is provided in~\ref{sec:appendix:questionnaire}.

\begin{table*}[!ht]
\small
\centering
\caption{Scales used in the survey study to measure stakeholder satisfaction, number of items, and reliability (Cronbach's Alpha) based on stakeholder-level response data ($N=1,746$)}
\label{tab:scalesstakeholdersatisfaction}
\begin{tabularx}{\textwidth}{@{}lXll@{}}
\toprule
\textbf{Construct variable} & \textbf{\# Items} & \textbf{Alpha} \\ 
\midrule
Satisfaction with value & 5 & .929 \\
Satisfaction with responsiveness & 4 & .826 \\
Satisfaction with release frequency & 3 & .896 \\
\bottomrule
\end{tabularx}
\end{table*}

\textbf{Control variables}: We included two control variables in the analyses. The first was the \textit{experience of teams with Agile}. This was operationalized with a single Likert question (1-7) ``I consider this team to be very experienced with Scrum/Agile.''. The second control variable was \textit{organization size}. It was operationalized with an ordinal variable that asked the initiator for each team to select the appropriate category: ``Between 1 And 50 Employees'', ``Between 51 And 500 Employees'', ``Between 501 And 5000 Employees'' and ``More Than 5000 Employees''

\subsection{Analysis}
\label{sec:analysis}
In this section, we describe the methods we employed to test our hypotheses. We used a combination of One-Way Analysis of Variance (ANOVA) and linear regression analyses to test for group differences between scaling approaches. While Analysis of Variance (ANOVA) is useful for identifying significant differences between groups, regression analyses also allowed us to control for the experience of teams with Agile and the size of the organization. 

The variables in our study were measured at the individual level and summarized to team-level mean averages. Such aggregation is only reasonable when sufficient variance exists at the group level, not just between individuals. We calculated the Intraclass Correlation (ICC)~\cite{hair2019multivariate} to determine the proportion of variance at the team level compared to the total variance. The ICC ranged between 35\% and 50\% for our independent variables, which exceeded the required threshold of 10\% suggested by Hair et al.~\cite{hair2019multivariate}. 

Since no data was missing, we did not deploy strategies to deal with missing data.

The normality of the distributions was assessed by inspecting skewness and kurtosis, Q-Q plots, and by performing a Kolmogorov-Smirnov test. The kurtosis ($<3$) and skewness ($<2$) remained below their recommended thresholds in the literature~\cite{de2014applications,hair2019multivariate}. However, a one-sample Kolmogorov-Smirnov test was significant for some variables which means their distributions deviate from a normal distribution. Further visual inspection of the Q-Q plots showed a more Cauchy-shaped distribution for these variables where the tails of the distribution are heavier and there is a higher propensity for extreme scores on both ends. Although ANOVA and regression analysis are generally robust against modest violations of normality~\cite{wilcox2011introduction}, we still opted for bootstrapping in our further analyses to normalize the distributions and reduce bias in our estimates~\cite{efron1992bootstrap}.

Afterward, we assessed the equality of variance between groups, which is an assumption for ANOVA. Levene's test was not significant for any of our variables. We employed Welch's ANOVA with Games-Howell post hoc tests throughout this paper as it does not assume equal variance while retaining similar statistical power~\cite{brown1974robust}.

Next, we tested the assumption of linearity. For a reliable interpretation of regression analyses, any increase in the independent variable must result in a consistent increase in the dependent variable. We performed curve fitting to assess if the relationship was significantly similar to linear, which was indeed the case. Another assumption of linear regression is homoscedasticity, meaning that the dependent variable has similar levels of variance across different levels of the independent variable~\cite{hair2019multivariate}. Violating this assumption greatly reduces statistical power~\cite{hair2019multivariate}. Thus, we assessed homoscedasticity by inspecting nine scatter plots for all pairs of continuous independent and dependent variables for inconsistent patterns but found none. Finally, multicollinearity was assessed by entering all independent variables one by one into a linear regression~\cite{gaskin2012data}. The Variance Inflation Factor (VIF) ranged between 1.023 and 1.348 and below the critical threshold of 10~\cite{hair2019multivariate} for all measures.

We performed a posthoc power analysis using G*Power~\cite{faul2009statistical}, version 3.1.9. We determined that our sample size allows us to correctly capture small effects ($f=.02$) with a statistical power of ~100\% ($1-\beta= 1.00$) for the sample of 4,013 teams. For the 529 teams with stakeholder evaluations, the statistical power for our regressions was also ~95\% ($1-\beta= 0.94$). So we are confident that our samples are large enough to provide a reliable outcome.

Because the chosen scaling approach of a team is a categorical variable, we created a dummy variable for each of the five scaling categories to indicate its use (1) or not (0). The baseline consists of all teams that do not use a scaling approach and all dummy variables are 0. For the five core processes of team effectiveness and the three indicators of stakeholder satisfaction, we performed eight controlled regression analyses with each indicator as the dependent variable, the dummy variables for scaling approaches, and the control variables (team experience and organization size) as independent variables. A second model was run for each regression analysis that did not include control variables to determine the difference in explained variance by the control variables. This illustrates the amount of variance explained by the scaling approaches alone and models that also consider other variables. For brevity, Section~\ref{sec:results} only reports the results from the controlled model and notes the variance explained by the uncontrolled model. Detailed results for both models are available in Appendix~\ref{sec:appendix:regressionanalysis}.

We report effect sizes throughout this study in addition to their significance. The substantial size of our sample can lead to type I errors where even a very small mean difference is statistically significant even though it is not meaningful in practice~\cite{cohen2013statistical}, which becomes more pronounced as sample size increases. This is one important contributor to the replication crisis that emerged in academic fields that relied on significance testing~\cite{ioannidis2005most}. The use of effect sizes requires researchers to also interpret the meaningfulness of the degree to which the results diverge from expectations~\cite{vacha2004estimate,kelley2012effect,russo2021pls}. 
For the Analysis of Variance, we calculate the eta-squared ($\eta^2$) as described in Ellis~\cite{ellis2010essential}. This statistic captures the amount of variance explained in the dependent variable by the independent variables. Values of less than or equal to .01, .06, and .14 are respectively considered to be small, moderate, and large in magnitude. Values below .01 indicate no effect. A 90\% confidence interval was calculated for the effect sizes based on the approach outlined by Smithson~\cite{smithson2003confidence}. For the linear regressions, we calculate the effect size with Cohen's $f^2$~\cite{cohen2013statistical} as described in Ellis~\cite{ellis2010essential}. This measure quantifies the proportion of variance in the dependent variable accounted for by the independent variable(s) in the regression model. Values of less than or equal to .02, .15, and .35 are respectively considered to be small, moderate, and large in magnitude. Values below .02 indicate no effect. The effect size for the full regression model is based on the explained variance ($R^2$), whereas the effect size of individual independent variables is based on the square of their part correlations ($R^2part$)~\cite{cohen2013applied}. A 90\% confidence interval was calculated for the effect sizes based on the approach outlined by Olkin \& Finn~\cite{olkin1995correlations}.

\section{Results}
\label{sec:results}
In this section, we report the results of our investigation. The first half of this section covers team effectiveness and the quality of core processes that determine it as reported by teams, and how they are influenced by the scaling approach. In the second half of this section, we investigate how the scaling approach influences complementary types of satisfaction as reported by stakeholders. Sample descriptives are shown in Table~\ref{tab:meansdeviationscorrelations}.

\begin{sidewaystable}
\centering
\caption{N, Means, Standard Deviations, Skewness, Kurtosis and Correlations (Pearson) for continuous variables. All reported correlations are statistically significant at $p <.01$}.
\label{tab:meansdeviationscorrelations}
\resizebox{\textwidth}{!}{%
\begin{tabular}{@{}Xlllllllllllllllllll@{}}
\toprule
 & \textbf{Variable} & \textbf{N} & \textbf{Mean} & \textbf{SD} & \textbf{Skewness} & \textbf{Kurtosis} & \textbf{1} & \textbf{2} & \textbf{3} & \textbf{4} & \textbf{5} & \textbf{6} & \textbf{7} & \textbf{8} & \textbf{9} & \textbf{10} \\ \midrule
1 & Responsiveness & 4,013 & 4.499 & 1.061 & -.486 & .025 & 1.000 &  &  &  &  &  &  &  &  &  \\
2 & Stakeholder Concern & 4,013 & 4.029 & 1.152 & -.178 & -.386 & .587 & 1.000 &  &  &  &  &  &  &  &  \\
3 & Continuous Improvement & 4,013 & 4.646 & .971 & -.544 & .197 & .642 & .733 & 1.000 &  &  &  &  &  &  &  \\
4 & Team Autonomy & 4,013 & 5.231 & .942 & -.888 & 1.068 & .523 & .521 & .638 & 1.000 &  &  &  &  &  &  \\
5 & Management Support & 4,013 & 4.792 & 1.473 & -.663 & -.155 & .461 & .554 & .589 & .497 & 1.000 &  &  &  &  &  \\
6 & Team Effectiveness & 4,013 & 5.152 & 1.060 & -.863 & .850 & .569 & .674 & .705 & .614 & .565 & 1.000 &  &  &  &  \\
7 & Stakeholder Satisfaction With Value & 549 & 5.469 & .933 & -.796 & .419 & .289 & .308 & .332 & .319 & .267 & .435 & 1.000 &  &  &  \\
8 & Stakeholder Satisfaction With Responsiveness & 549 & 5.463 & .920 & -.605 & .587 & .272 & .300 & .272 & .272 & .270 & .327 & .661 & 1.000 &  &  \\
9 & Stakeholder Satisfaction With Release Frequency & 549 & 5.208 & 1.098 & -.643 & .301 & .287 & .240 & .301 & .291 & .250 & .372 & .766 & .561 & 1.000 &  \\
\multicolumn{17}{c}{\emph{Control Variables}} \\ \addlinespace
10 & Team Experience With Agile & 4,013 & 4.496 & 1.518 & -.544 & .039 & .589 & .612 & .687 & .556 & .537 & .616 & .281 & .228 & .260 & 1.000 \\
\bottomrule
\end{tabular}
}
\end{sidewaystable}

\subsection{Team effectiveness by scaling approach}
\begin{figure}[htb]
\centering
\includegraphics[height=2.5in]{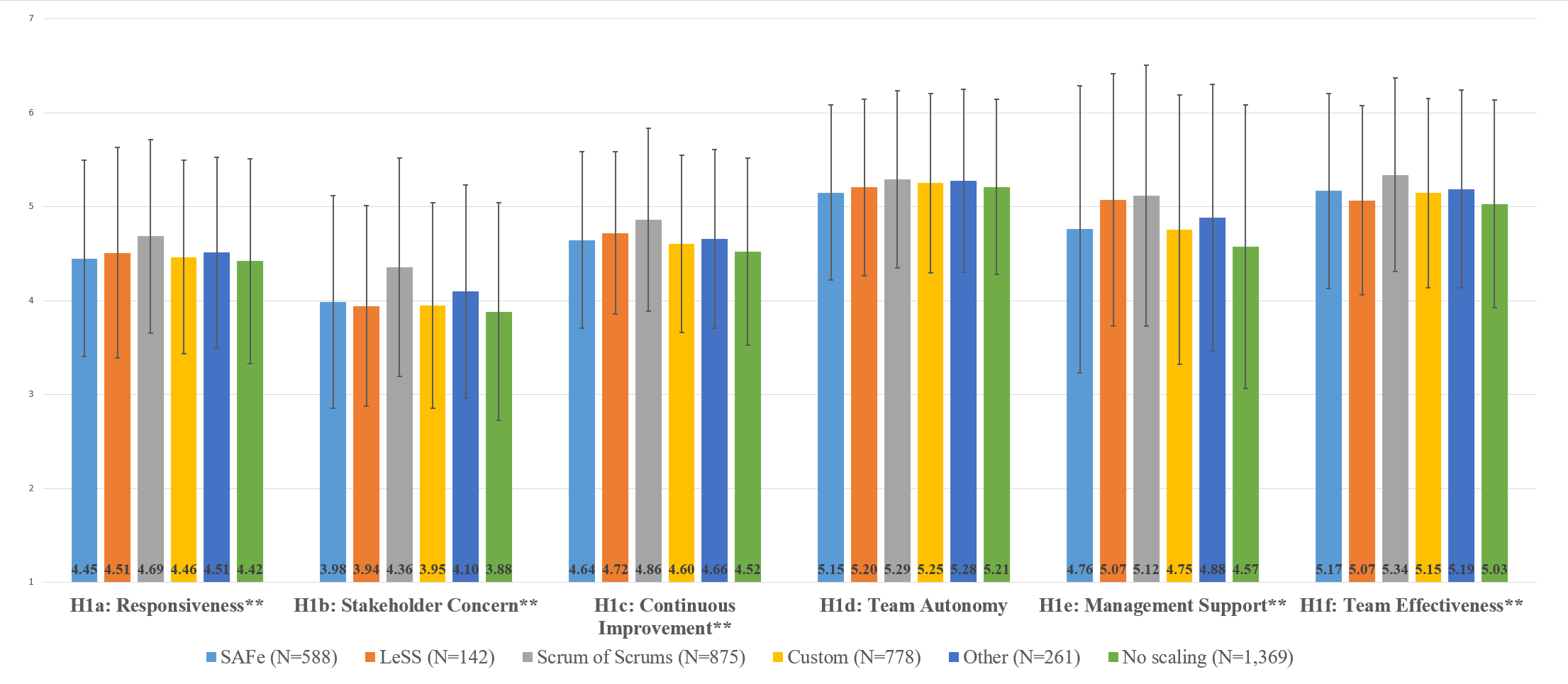}
\caption{Means for the five core indicators of Agile team effectiveness and effectiveness compared by scaling approach for 15,078 team members aggregated into 4,013 teams. The bars represent 1 standard deviation. All factors score significantly different between groups at $p < .05$. }
\label{fig:teameffectivenessbyscalingframework}
\end{figure}

We begin with the results of the Analysis of Variance (ANOVA) that compared the indicators of team effectiveness by scaling approach. Table~\ref{tab:teameffectivenessbyscalingframework} shows the means, standard deviations, analysis of variance, and effect sizes compared by the scaling framework in use. Figure~\ref{fig:teameffectivenessbyscalingframework} presents the results in visual form.

\begin{sidewaystable}
\centering
\caption{Means, Standard Deviations, one-way Analyses of Variance (Welsh) and effect size ($\eta^2$) with 90\% confidence interval for core indicators of Scrum Team Effectiveness compared by scaling framework for 11.376 team members aggregated into 3.102 teams. *: statistically significant at $p <.05$. **: statistically significant at $p <.01$}
\label{tab:teameffectivenessbyscalingframework}
\resizebox{\textwidth}{!}{%
\begin{tabular}{llllllllllllllll}
\toprule
Indicator & \multicolumn{2}{l}{SAFe (N=588)} & \multicolumn{2}{l}{LeSS (N=142)} & \multicolumn{2}{l}{Scrum of Scrums (N=875)} & \multicolumn{2}{l}{Custom (N=778)} & \multicolumn{2}{l}{Other (N=261)} & \multicolumn{2}{l}{No scaling (N=1,369)} & F(5,3101) & P & $\eta^2$ / 90\% CI \\
 & M & SD & M & SD & M & SD & M & SD & M & SD & M & SD &  &  &  \\
\midrule
H1a: Responsiveness** & 4.448 & 1.045 & 4.509 & 1.120 & 4.685 & 1.029 & 4.463 & 1.032 & 4.511 & 1.015 & 4.419 & 1.093 & 7.456 & .000 & .009 {[}.004, .014{]} \\
H1b: Stakeholder Concern** & 3.985 & 1.133 & 3.943 & 1.066 & 4.356 & 1.164 & 3.945 & 1.094 & 4.096 & 1.138 & 3.882 & 1.157 & 20.338 & .000 & .025 {[}.016, .032{]} \\
H1c: Continuous Improvement** & 4.645 & .937 & 4.718 & .864 & 4.862 & .971 & 4.605 & .943 & 4.657 & .948 & 4.521 & .994 & 13.841 & .000 & .017 {[}.010, .023{]} \\
H1d: Team Autonomy & 5.151 & .929 & 5.204 & .938 & 5.294 & .943 & 5.252 & .955 & 5.277 & .972 & 5.208 & .932 & 2.020 & .073 & .003 {[}.000, .005{]} \\
H1e: Management Support** & 4.760 & 1.527 & 5.071 & 1.343 & 5.115 & 1.387 & 4.754 & 1.436 & 4.883 & 1.417 & 4.574 & 1.506 & 16.093 & .000 & .020 {[}.012, .026{]} \\
H1f: Team Effectiveness** & 5.169 & 1.039 & 5.068 & 1.007 & 5.340 & 1.030 & 5.147 & 1.008 & 5.187 & 1.053 & 5.031 & 1.107 & 9.460 & .000 & .012 {[}.006, .017{]} \\
\bottomrule
\end{tabular}
}
\end{sidewaystable}

Our results show that \textbf{team autonomy is not significantly different between scaling approaches} (H1d). However, team effectiveness and all other core processes are significantly different between scaling frameworks (H1a-c \& H1e-f, $p < .05$). This means that the level of responsiveness, stakeholder concern, continuous improvement, management support, and team effectiveness itself varies between scaling approaches, but the level of experienced autonomy by teams did not. However, the size of the observed effects ($\eta^2$) is qualified as \textit{small}. Of the scaling approaches under investigation, ``Scrum of Scrums'' scores the highest for all processes. The results for the other scaling approaches are comparable across indicators, with very small differences, except for the factor ``Management Support'' where both ``Scrum of Scrums'' ($M=5.11$) and ``LeSS'' ($M=5.07$) score higher than the other approaches ($M=4.57$ and $M=4.88$). Thus, the simpler approaches to scaling seem to have a slight edge over more complex approaches.

In the following subsections, we explore how each indicator is influenced by the scaling approach while controlling for team experience and organization size.

\subsubsection{Does the scaling approach influence the responsiveness of teams?}
We performed regression analysis to predict the responsiveness of teams based on the scaling approach in use and while controlling for the experience of teams with Agile and organization size (see Table~\ref{tab:responsivenessbyscalingapproach}). The regression model was significant, $F(7,3788) = 293.876, p < .01$, and explained 35.1\% of the observed variance in responsiveness compared to 1.0\% by a secondary regression without control variables (see Section~\ref{sec:analysis} for the rationale behind the two analyses). The overall effect size of the model is qualified as \textit{large}, $f^2 = .545, 90\% CI [.519, .567]$. 

Of the scaling approaches, ``SAFe'' has a significant effect on responsiveness, although its effect size is qualified as \textit{none}, $beta = -.042, p < .01, f^2 = .001, 90\% CI [-.001, .004]$. ``Custom approach'' also has a smaller negative but significant effect on responsiveness, but its effect size is also qualified as \textit{none}, $beta = -.032, p < .05, f^2 = .001, 90\% CI [-.001, .002]$. So while some positive or negative effects are observed of the studied scaling approaches, these effects are so small as to be practically irrelevant.

Of the control variables, the experience of teams with Agile has a significant effect that is qualified as \textit{large} based on its effect size, $beta = .594, p < .01, f^2 = .521, 90\% CI [.497, .545]$. Organization size has a significant effect, but its effect size is qualified as \textit{none}, $beta -.040, p < .001, f^2 = .001, 90\% CI [-.001, .004]$. Thus, experienced teams are clearly more responsive than less-experienced teams, regardless of what scaling approach is used. But teams are not more or less responsive between differently sized organizations.

\begin{table}[!ht]
\centering
\small
\caption{Coefficients, Standard Errors (SE), Beta's, t-values, significance, explained variance ($R^2part$) and effect size ($f^2$) with 90\% confidence intervals for the chosen scaling approach, experience of teams with Agile, and the size of the organization on team responsiveness. *: statistically significant at $p <.05$. **: statistically significant at $p <.01$}
\label{tab:responsivenessbyscalingapproach}
\begin{tabular}{@{}m{3cm}m{1cm}m{1cm}m{1cm}m{1cm}m{1cm}m{1cm}m{2.5cm}@{}}
\toprule
\textbf{Variable} & \textbf{Coeff.} & \textbf{SE} & \textbf{$\beta$} & \textbf{t} & \textbf{$p$} & \textbf{$R^2$} & \textbf{$f^2$ / 90\% CI} \\
\midrule
Intercept** & 2.799 & .058 & .000 & 48.297 & .000 &  &  \\ \addlinespace
Uses SAFe** & -.124 & .045 & -.042 & -2.777 & .006 & .001 & .001 [-.001, .004] \\ \addlinespace
Uses LESS & -.042 & .077 & -.007 & -.545 & .586 & .000 & .000 [.000, .000] \\ \addlinespace
Uses Scrum of Scrums & .003 & .039 & .001 & .066 & .947 & .000 & .000 [.000, .000] \\ \addlinespace
Custom approach* & -.084 & .040 & -.032 & -2.113 & .035 & .001 & .001 [-.001, .002] \\ \addlinespace
Other approach & -.045 & .060 & -.010 & -.751 & .453 & .000 & .000 [-.001, .001] \\ \addlinespace
Control: Team Experience with Agile** & .415 & .009 & .594 & 44.749 & .000 & .343 & .521 [.497, .545] \\ \addlinespace
Control: Size of organization** & -.044 & .015 & -.040 & -2.930 & .003 & .001 & .001 [-.001, .004] \\
\bottomrule
\end{tabular}
\vspace{2em}
\end{table}

\subsubsection{Does the scaling approach influence stakeholder concern of teams?}
We performed regression analysis to predict stakeholder concern of teams based on the scaling approach in use (see Table~\ref{tab:stakeholderconcernbyscalingapproach}). The regression model was significant, $F(7,3788) = 340.405, p < .01$, and explained 38.5\% of the observed variance in responsiveness compared to 2.5\% by a secondary regression without control variables. The overall effect size of the model is qualified as \textit{large}, $f^2 = .629, 90\% CI [.600, .658]$. 

Of the scaling approaches, ``SAFe'' has a significant negative effect on stakeholder concern, although its effect size is qualified as \textit{none}, $beta = -.046, p < .01, f^2 = .002, 90\% CI [-.001, .005]$. Stakeholder concern is also positively and significantly affected by ``Scrum of Scrums'', but its effect size is also qualified as \textit{none}, $beta = .057, p < .01, f^2 = .001, 90\% CI [-.001, .003]$. ``Custom approach'' has a significant negative effect on stakeholder concern that is qualified as \textit{none} based on its size, $beta = -.029, p < .01, f^2 = .001, 90\% CI [-.001, .003]$. So while some positive or negative effects are observed of the studied scaling approaches, these effects are so small as to be practically irrelevant.

Of the control variables, the experience of teams with Agile has a significant effect that is qualified as \textit{large} based on its size, $beta = .604, p < .01, f^2 = .549, 90\% CI [.519, .578]$. Organization size has a significant effect, but its effect size is qualified as \textit{none}, $beta .057, p < .001, f^2 = .003, 90\% CI [-.001, .007]$. The results show that experienced teams are much more focused on the needs of their stakeholders, regardless of the scaling approach in use. But teams are not more or less concerned with their stakeholders between differently sized organizations. 

\begin{table}[!ht]
\centering
\small
\caption{Coefficients, Standard Errors (SE), Beta's, t-values, significance, explained variance ($R^2$) and effect size ($f^2$) with 90\% confidence intervals for the chosen scaling approach, experience of teams with Agile, and the size of the organization on stakeholder concern. *: statistically significant at $p <.05$. **: statistically significant at $p <.01$}
\label{tab:stakeholderconcernbyscalingapproach}
\begin{tabular}{@{}m{3cm}m{1cm}m{1cm}m{1cm}m{1cm}m{1cm}m{1cm}m{2.5cm}@{}}
\toprule
\textbf{Variable} & \textbf{Coeff.} & \textbf{SE} & \textbf{$\beta$} & \textbf{t} & \textbf{$p$} & \textbf{$R^2$} & \textbf{$f^2$ / 90\% CI} \\
\midrule
Intercept** & 1.788 & .061 & .000 & 29.290 & .000 &  &  \\ \addlinespace
Uses SAFe** & -.148 & .047 & -.046 & -3.127 & .002 & .002 & .002 [-.001, .005] \\ \addlinespace
Uses LESS & -.110 & .081 & -.018 & -1.356 & .175 & .000 & .000 [-.001, .002] \\ \addlinespace
Uses Scrum of Scrums** & .160 & .041 & .057 & 3.879 & .000 & .002 & .002 [-.001, .006] \\ \addlinespace
Custom approach* & -.083 & .042 & -.029 & -1.991 & .047 & .001 & .001 [-.001, .003] \\ \addlinespace
Other approach & .025 & .063 & .005 & .395 & .693 & .000 & .000 [.000, .000] \\ \addlinespace
Control: Team Experience with Agile** & .457 & .010 & .604 & 46.765 & .000 & .354 & .549 [.519, .578] \\ \addlinespace
Control: Size of organization** & .070 & .016 & .057 & 4.377 & .000 & .003 & .003 [-.001, .007] \\
\bottomrule
\end{tabular}
\vspace{2em}
\end{table}

\subsubsection{Does the scaling approach influence continuous improvement in teams?}
We performed regression analysis to predict the level of continuous improvement in teams based on the scaling approach in use (see Table~\ref{tab:continuousimprovementbyscalingapproach}). The regression model was significant, $F(7,3788) = 488.946, p < .01$, and explained 47.4\% of the observed variance in responsiveness compared to 1.7\% by a secondary regression without control variables. The overall effect size of the model is qualified as \textit{large}, $f^2 = .900, 90\% CI [.877, .928]$. 

None of the scaling approaches significantly predict the level of continuous improvement in teams. This means that teams can engage in continuous improvement equally between different scaling approaches.

Of the control variables, only the experience of teams with Agile has a significant effect that is qualified as \textit{large} based on its size, $beta = .686, p < .01, f^2 = .846, 90\% CI [.817, .874]$. Thus, experienced teams are clearly better able to engage in continuous improvement than less-experienced teams, regardless of what scaling approach is used. However, organization size does not meaningfully influence the level of continuous improvement in a team.

\begin{table}[!ht]
\centering
\small
\caption{Coefficients, Standard Errors (SE), Beta's, t-values, significance, explained variance ($R^2$) and effect size ($f^2$) with 90\% confidence intervals for the chosen scaling approach, experience of teams with Agile, and the size of the organization on continuous improvement in teams. *: statistically significant at $p <.05$. **: statistically significant at $p <.01$}
\label{tab:continuousimprovementbyscalingapproach}
\begin{tabular}{@{}m{3cm}m{1cm}m{1cm}m{1cm}m{1cm}m{1cm}m{1cm}m{2.5cm}@{}}
\toprule
\textbf{Variable} & \textbf{Coeff.} & \textbf{SE} & \textbf{$\beta$} & \textbf{t} & \textbf{$p$} & \textbf{$R^2$} & \textbf{$f^2$ / 90\% CI} \\
\midrule
Intercept** & 2.728 & .048 & .000 & 57.274 & .000 &  &  \\ \addlinespace
Uses SAFe & -.058 & .037 & -.021 & -1.578 & .115 & .000 & .000 [-.001, .002] \\ \addlinespace
Uses LESS & .042 & .063 & .008 & .670 & .503 & .000 & .000 [-.001, .001] \\ \addlinespace
Uses Scrum of Scrums & .052 & .032 & .022 & 1.630 & .103 & .000 & .000 [-.001, .002] \\ \addlinespace
Custom approach & -.047 & .033 & -.019 & -1.442 & .149 & .000 & .000 [-.001, .002] \\ \addlinespace
Other approach & -.012 & .049 & -.003 & -.250 & .802 & .000 & .000 [.000, .000] \\ \addlinespace
Control: Team Experience with Agile** & .438 & .008 & .686 & 57.478 & .000 & .458 & .846 [.817, .874] \\ \addlinespace
Control: Size of organization & -.017 & .012 & -.017 & -1.385 & .166 & .000 & .000 [-.001, .002] \\
\bottomrule
\end{tabular}
\vspace{2em}
\end{table}

\subsubsection{Does the scaling approach influence team autonomy?}
We performed regression analysis to predict team autonomy based on the scaling approach in use (see Table~\ref{tab:teamautonomybyscalingapproach}). The regression model was significant, $F(7,2958) = 254.352, p < .01$, and explained 32.0\% of the observed variance in responsiveness compared to 0.3\% by a secondary regression without control variables. The overall effect size of the model is qualified as \textit{large}, $f^2 = .470, 90\% CI [.441, .500]$. 

Of the scaling approaches, ``SAFe'' has a significant negative effect on team autonomy that is qualified as \textit{none} based on its size, $beta = -.070, p < .01, f^2 = .004, 90\% CI [-.001, .008]$. ``Scrum of Scrums'' also has a significant negative effect on team autonomy and is also qualified as \textit{none} based on its size, $beta = -.060, p < .01, f^2 = .003, 90\% CI [-.001, .007]$. The other scaling approaches do not affect team autonomy. Thus, any observed effects of scaling approaches are so small as to be practically irrelevant.

Of the control variables, the experience of teams with Agile has a significant effect that is qualified as \textit{large} based on its size, $beta = .571, p < .01, f^2 = .465, 90\% CI [.435, .494]$. Organization size has a significant negative effect with an effect size that is qualified as \textit{none}, $beta -.032, p < .05, f^2 = .001, 90\% CI [-.001, .003]$. Thus, experienced teams are more autonomous than less-experienced teams, regardless of what scaling approach is used. However, teams can be equally autonomous in small and large organizations.

\begin{table}[!ht]
\centering
\small
\caption{Coefficients, Standard Errors (SE), Beta's, t-values, significance, explained variance ($R^2$) and effect size ($f^2$) with 90\% confidence intervals for the chosen scaling approach, experience of teams with Agile, and the size of the organization on team autonomy. *: statistically significant at $p <.05$. **: statistically significant at $p <.01$}
\label{tab:teamautonomybyscalingapproach}
\begin{tabular}{@{}m{3cm}m{1cm}m{1cm}m{1cm}m{1cm}m{1cm}m{1cm}m{2.5cm}@{}}
\toprule
\textbf{Variable} & \textbf{Coeff.} & \textbf{SE} & \textbf{$\beta$} & \textbf{t} & \textbf{$p$} & \textbf{$R^2$} & \textbf{$f^2$ / 90\% CI} \\
\midrule
Intercept** & 3.813 & .053 & .000 & 72.545 & .000 &  &  \\ \addlinespace
Uses SAFe** & -.185 & .041 & -.070 & -4.561 & .000 & .004 & .004 [-.001, .008] \\ \addlinespace
Uses LESS & -.119 & .070 & -.024 & -1.714 & .087 & .001 & .001 [-.001, .002] \\ \addlinespace
Uses Scrum of Scrums** & -.137 & .035 & -.060 & -3.856 & .000 & .003 & .003 [-.001, .007] \\ \addlinespace
Custom approach & -.068 & .036 & -.029 & -1.882 & .060 & .001 & .001 [-.001, .003] \\ \addlinespace
Other approach & -.054 & .054 & -.014 & -.999 & .318 & .000 & .000 [-.001, .001] \\ \addlinespace
Control: Team Experience with Agile** & .353 & .008 & .571 & 42.028 & .000 & .317 & .465 [.435, .494] \\ \addlinespace
Control: Size of organization* & -.032 & .014 & -.032 & -2.332 & .020 & .001 & .001 [-.001, .003] \\
\bottomrule
\end{tabular}
\vspace{2em}
\end{table}

\subsubsection{Does the scaling approach influence management support?}
We performed regression analysis to predict management support based on the scaling approach in use (see Table~\ref{tab:managementsupportbyscalingapproach}). The regression model was significant, $F(7,3788) = 227.365, p < .01$, and explained 29.6\% of the observed variance in responsiveness compared to 2.0\% by a secondary regression without control variables. The overall effect size of the model is qualified as \textit{large}, $f^2 = .418, 90\% CI [.388, .447]$. 

Of the scaling approaches, ``LeSS'' has a significant and positive effect on management support that is qualified as \textit{none} based on its effect size $beta = .042, p < .01, f^2 = .002, 90\% CI [-.001, .005]$. Similarly, ``Scrum of Scrums'' also has a significant positive effect on management support that is also qualified as \textit{none} based on its size, $beta = .060, p < .01, f^2 = .003, 90\% CI [-.001, .007]$. The other scaling approaches do not affect management support. So while some positive effects are observed of the scaling approaches, these effects are so small as to be practically irrelevant.

Of the control variables, the experience of teams with Agile has a significant effect that is qualified as \textit{large} based on its size, $beta = .533, p < .01, f^2 = .381, 90\% CI [.352, .411]$. Organization size has a significant negative effect with an effect size that is qualified as \textit{none}, $beta -.054, p < .01, f^2 = .003, 90\% CI [-.001, .007]$. This means that experienced experience more support from management than less-experienced teams, regardless of what scaling approach is used. However, the level of management support is not different between smaller and larger organizations.

\begin{table}[!ht]
\centering
\small
\caption{Coefficients, Standard Errors (SE), Beta's, t-values, significance, explained variance ($R^2$) and effect size ($f^2$) with 90\% confidence intervals for the chosen scaling approach, experience of teams with Agile, and the size of the organization on management support. *: statistically significant at $p <.05$. **: statistically significant at $p <.01$}
\label{tab:managementsupportbyscalingapproach}
\begin{tabular}{@{}m{3cm}m{1cm}m{1cm}m{1cm}m{1cm}m{1cm}m{1cm}m{2.5cm}@{}}
\toprule
\textbf{Variable} & \textbf{Coeff.} & \textbf{SE} & \textbf{$\beta$} & \textbf{t} & \textbf{$p$} & \textbf{$R^2$} & \textbf{$f^2$ / 90\% CI} \\
\midrule
Intercept** & 2.633 & .084 & .000 & 31.430 & .000 &  &  \\ \addlinespace
Uses SAFe & .018 & .065 & .004 & .274 & .784 & .000 & .000 [-.001, .000] \\ \addlinespace
Uses LESS** & .332 & .111 & .042 & 2.994 & .003 & .002 & .002 [-.001, .005] \\ \addlinespace
Uses Scrum of Scrums** & .213 & .057 & .060 & 3.774 & .000 & .003 & .003 [-.001, .007] \\ \addlinespace
Custom approach & .044 & .057 & .012 & .771 & .440 & .000 & .000 [-.001, .001] \\ \addlinespace
Other approach & .166 & .086 & .028 & 1.931 & .054 & .001 & .001 [-.001, .003] \\ \addlinespace
Control: Team Experience with Agile** & .517 & .013 & .533 & 38.540 & .000 & .276 & .381 [.352, .411] \\ \addlinespace
Control: Size of organization** & -.089 & .022 & -.057 & -4.048 & .000 & .003 & .003 [-.001, .007] \\
\bottomrule
\end{tabular}
\vspace{2em}
\end{table}

\subsubsection{Does the scaling approach influence overall team effectiveness?}
Finally, we performed regression analysis to predict team effectiveness based on the scaling approach in use (see Table~\ref{tab:teameffectivenessbyscalingapproach}). The regression model was significant, $F(7,3788) = 337.576, p < .01$, and explained 38.4\% of the observed variance in responsiveness compared to 1.2\% by a secondary regression without control variables. The overall effect size of the model is qualified as \textit{large}, $f^2 = .624, 90\% CI [.570, .619]$. 

None of the scaling approaches significantly affect team effectiveness. This means that teams can be equally effective regardless of the scaling approach in use.

Of the control variables, only the experience of teams with Agile has a significant effect that is qualified as \textit{large} based on its size, $beta = .618, p < .01, f^2 = .591, 90\% CI [.562, .620]$. Thus, teams are clearly more effective when they are more experienced. Teams can also be equally effective in smaller and larger organizations.

\begin{table}[!ht]
\centering
\small
\caption{Coefficients, Standard Errors (SE), Beta's, t-values, significance, explained variance ($R^2$) and effect size ($f^2$) with 90\% confidence intervals for the chosen scaling approach, experience of teams with Agile, and the size of the organization on team effectiveness. *: statistically significant at $p <.05$. **: statistically significant at $p <.01$}
\label{tab:teameffectivenessbyscalingapproach}
\begin{tabular}{@{}m{3cm}m{1cm}m{1cm}m{1cm}m{1cm}m{1cm}m{1cm}m{2.5cm}@{}}
\toprule
\textbf{Variable} & \textbf{Coeff.} & \textbf{SE} & \textbf{$\beta$} & \textbf{t} & \textbf{$p$} & \textbf{$R^2$} & \textbf{$f^2$ / 90\% CI} \\
\midrule
Intercept** & 3.205 & .056 & .000 & 57.001 & .000 &  &  \\ \addlinespace
Uses SAFe & -.060 & .043 & -.020 & -1.369 & .171 & .000 & .000 [-.001, .002] \\ \addlinespace
Uses LESS & -.093 & .075 & -.016 & -1.243 & .214 & .000 & .000 [-.001, .001] \\ \addlinespace
Uses Scrum of Scrums & .025 & .038 & .010 & .659 & .510 & .000 & .000 [-.001, .001] \\ \addlinespace
Custom approach & -.019 & .039 & -.007 & -.505 & .614 & .000 & .000 [.000, .001] \\ \addlinespace
Other approach & -.012 & .058 & -.003 & -.212 & .832 & .000 & .000 [.000, .000] \\ \addlinespace
Control: Team Experience with Agile** & .430 & .009 & .618 & 47.802 & .000 & .371 & .591 [.562, .620] \\ \addlinespace
Control: Size of organization & .011 & .015 & .009 & .719 & .472 & .000 & .000 [-.001, .001] \\
\bottomrule
\end{tabular}
\vspace{2em}
\end{table}

\subsubsection{Summary for team effectiveness and core processes}
A summary of the beta's (i.e., indicating the direction and strength of the relationships) and effect sizes ($f^2$) that were identified in the regression analyses in this section is provided in Table~\ref{tab:teameffectivenesssummary} along with their significance and effect size classifications~\cite{cohen2013statistical}. After controlling for team experience and organization size, \textbf{we found some significant effects for ``SAFe'', ``Scrum of Scrums'', ``LeSS'' and ``Custom approach'' on the indicators of team effectiveness ($H1a-H1f$), although their size was so small as to be qualified as \textit{none}}. However, a large effect was found for the control variable team experience on all indicators. 

This concludes the results for the indicators of team effectiveness as reported by teams. We now turn to the satisfaction reported by the stakeholders of those teams.

\begin{sidewaystable}
\centering
\caption{Summary of beta's and effect sizes ($f^2$) by indicators of team effectiveness for scaling approaches and control variables along with significance and effect size classification. *: statistically significant at $p <.05$. **: statistically significant at $p <.01$. a: no effect ($f^2 < .02$), b: small effect ($f^2 >= .02$), c: moderate effect ($f^2 >= .15$), d: large effect ($f^2 >= .35$)}
\label{tab:teameffectivenesssummary}
\resizebox{\textwidth}{!}{%
\begin{tabular}{lllllllllllll}
\toprule
Variable & \multicolumn{2}{l}{Responsiveness} & \multicolumn{2}{l}{Stakeholder Concern} & \multicolumn{2}{l}{Continuous Improvement} & \multicolumn{2}{l}{Team Autonomy} & \multicolumn{2}{l}{Management Support} & \multicolumn{2}{l}{Team Effectiveness} \\
& $\beta$ & $f^2$ & $\beta$ & $f^2$ & $\beta$ & $f^2$ & $\beta$ & $f^2$ & $\beta$ & $f^2$ & $\beta$ & $f^2$ \\
\midrule
SAFe & -.042** & $.001^a$ & -.046** & $.002^a$ & -.021 & $.000^a$ & -.070** & $.004^a$ & .004 & $.000^a$ & -.020 & $.000^a$ \\
LeSS & -.007 & $.000^a$ & -.018 & $.000^a$ & .008 & $.000^a$ & -.024 & $.001^a$ & .042** & $.002^a$ & -.016 & $.000^a$ \\
Scrum of Scrums & .001 & $.000^a$ & .057** & $.002^a$ & .022 & $.000^a$ & -.060** & $.003^a$ & .060** & $.003^a$ & .010 & $.000^a$ \\
Custom approach & -.032* & $.001^a$ & -.029* & $.001^a$ & -.019 & $.000^a$ & -.029 & $.001^a$ & .012 & $.000^a$ & -.007 & $.000^a$ \\
Other approach & -.010 & $.000^a$ & .005 & $.000^a$ & -.003 & $.000^a$ & -.014 & $.000^a$ & .028 & $.001^a$ & -.003 & $.000^a$ \\
Control: Team Experience with Agile & .594** & $.521^d$ & .604** & $.549^d$ & .686** & $.846^d$ & .571** & $.465^d$ & .533** & $.381^d$ & .618** & $.591^d$ \\
Control: Size of organization & -.040** & $.001^a$ & .057** & $.003^a$ & -.017 & $.000^a$ & -.032* & $.001^a$ & -.057** & $.003^a$ & .009 & $.000^a$ \\
\bottomrule
\end{tabular}
}
\end{sidewaystable}

\subsection{Stakeholder satisfaction by scaling approach}\
Here, we explore the satisfaction reported by stakeholders with the responsiveness, release frequency, and value delivered by their teams. Evaluations were collected from 1,841 stakeholders for 529 of the teams used in this study.

\begin{figure}[htb]
\centering
\includegraphics[height=2in]{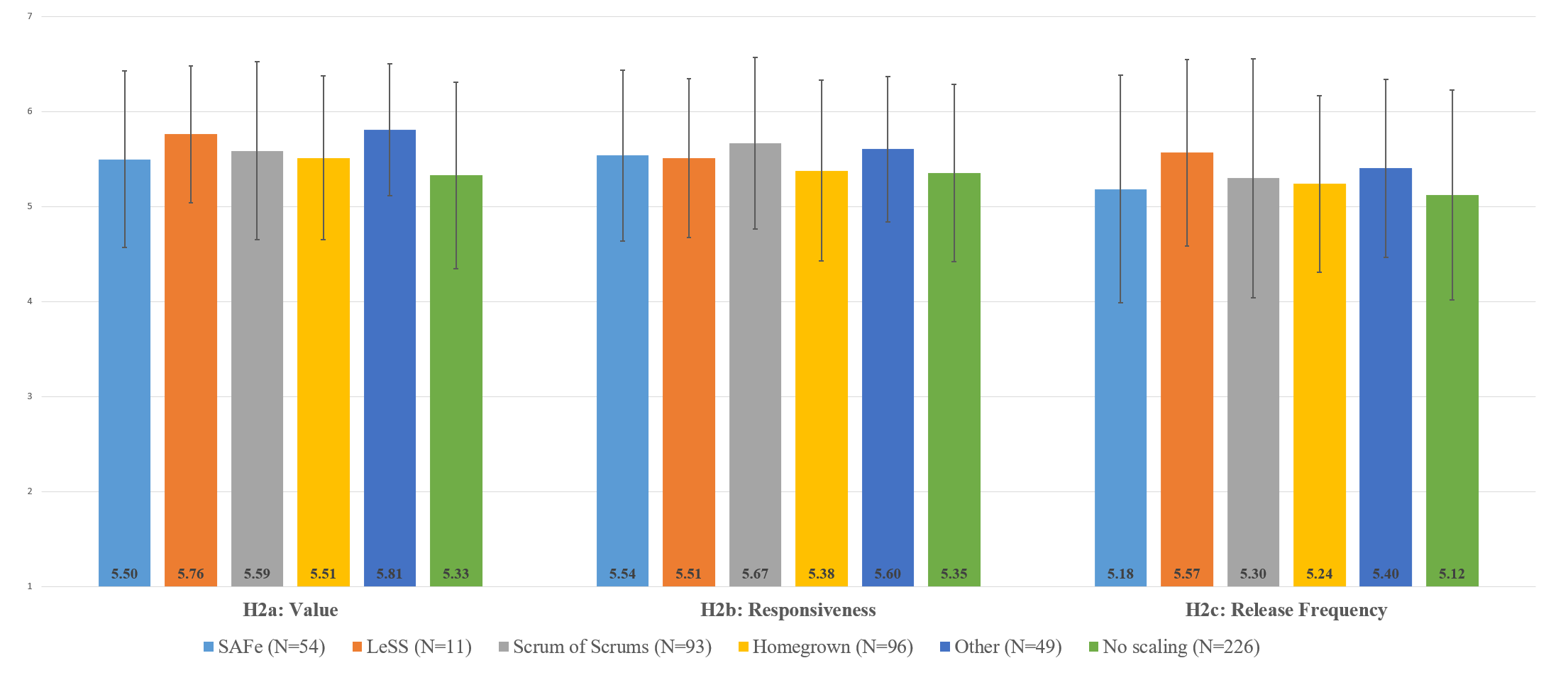}
\caption{Means for the four indicators of stakeholder satisfaction compared by scaling approach. The bars represent 1 standard deviation. Factors marked with ** are significantly different between groups at $p < .01$. Factors marked with * are significantly different between groups at $ < .05$.}
\label{fig:stakeholderevaluationbyscalingapproach}
\end{figure}

We begin with the results of an Analysis of Variance (ANOVA) that compared the indicators of team effectiveness by scaling approach. Table~\ref{tab:stakeholderevaluationbyscalingapproach} shows the means, standard deviations, analysis of variance, and effect sizes compared by the scaling framework in use. Figure~\ref{fig:stakeholderevaluationbyscalingapproach} presents the results in visual form. \textbf{A significant difference exists for the value delivered (H2a, $p < .01$), but not for responsiveness (H2b) and release frequency (H2c). The effect size for this difference is qualified as \textit{small}. For satisfaction with value, ``LeSS'' and ``Other approach'' score relatively highest.}

\begin{sidewaystable}
\centering
\caption{Means, Standard Deviations, one-way Analyses of Variance (Welsh) and effect size ($\eta^2$) for indicators of stakeholder satisfaction compared by scaling approach. *: statistically significant at $p <.05$. **: statistically significant at $p <.01$}
\label{tab:stakeholderevaluationbyscalingapproach}
\resizebox{\textwidth}{!}{%
\begin{tabular}{llllllllllllllll}
\toprule
Indicator & \multicolumn{2}{l}{SAFe (N=54)} & \multicolumn{2}{l}{LeSS (N=11)} & \multicolumn{2}{l}{Scrum of Scrums (N=93)} & \multicolumn{2}{l}{Homegrown (N=96)} & \multicolumn{2}{l}{Other (N=49)} & \multicolumn{2}{l}{No scaling (N=226)} & F(5,528) & P & $\eta^2$ / 90\% CI \\
 & M & SD & M & SD & M & SD & M & SD & M & SD & M & SD &  &  &  \\
\midrule
H2a: Value & 5.497 & .929 & 5.760 & .719 & 5.587 & .937 & 5.512 & .862 & 5.807 & .691 & 5.327 & .980 & 2.978 & .012 & .028 {[}.003, .046{]} \\
H2b: Responsiveness & 5.535 & .900 & 5.507 & .836 & 5.667 & .905 & 5.378 & .953 & 5.604 & .765 & 5.352 & .933 & 2.077 & .067 & .019 {[}.000, .034{]} \\
H2c: Release Frequency & 5.182 & 1.197 & 5.567 & .979 & 5.299 & 1.257 & 5.239 & .931 & 5.401 & .934 & 5.120 & 1.106 & .974 & .433 & .009 {[}.000, .017{]} \\
\bottomrule
\end{tabular}
}
\end{sidewaystable}

In the following subsections, we explore how the satisfaction of stakeholders with each area is influenced by the scaling approach with and without controlling for team experience and organization size.

\subsubsection{Does the scaling approach influence stakeholder satisfaction with value?}
To explore this question, we performed a linear regression analysis to predict the satisfaction with value as reported by stakeholders based on the scaling approach in use (see Table~\ref{tab:stakeholdersatisfactionwithvaluebyscalingapproach}). The regression model was significant, $F(7,521) = 8.582, p < .01$, and explained 10.3\% of the observed variance in responsiveness compared to 2.8\% by a secondary regression without control variables. The overall effect size of the model is qualified as \textit{small}, $f^2 = .101, 90\% CI [.057, .153]$. This means that the chosen scaling approach explains little of the observed variance in stakeholder satisfaction with value.

Of the scaling approaches, only ``Other approach'' has a significant and positive effect on satisfaction with value that is qualified as \textit{small} based on its effect size, $beta = .133, p < .01, f^2 = .016, 90\% CI [-.004, .040]$. Thus, the stakeholders of teams that use scaling approaches other than the ones we identified specifically in this study appear to be a little more satisfied with the value produced by teams, although the size of this effect is small.

Of the control variables, only the experience of teams with Agile has a significant effect that is qualified as \textit{small} based on its size, $beta = .284, p < .01, f^2 = .082, 90\% CI [.041, .130]$. This means that experienced teams are a little more able to satisfy their stakeholders, although this effect is small and determined more by other factors.

\begin{table}[!ht]
\centering
\small
\caption{Coefficients, Standard Errors (SE), Beta's, t-values, significance, explained variance ($R^2$) and effect size ($f^2$) with 90\% confidence intervals for the chosen scaling approach, experience of teams with Agile, and the size of the organization on stakeholder satisfaction with value. *: statistically significant at $p <.05$. **: statistically significant at $p <.01$}
\label{tab:stakeholdersatisfactionwithvaluebyscalingapproach}
\begin{tabular}{@{}m{3cm}m{1cm}m{1cm}m{1cm}m{1cm}m{1cm}m{1cm}m{2.5cm}@{}}
\toprule
\textbf{Variable} & \textbf{Coeff.} & \textbf{SE} & \textbf{$\beta$} & \textbf{t} & \textbf{$p$} & \textbf{$R^2$} & \textbf{$f^2$ / 90\% CI} \\
\midrule
Intercept** & 4.136 & .215 & .000 & 19.205 & .000 &  &  \\ \addlinespace
Uses SAFe & .032 & .141 & .011 & .230 & .818 & .000 & .000 [-.001, .002] \\ \addlinespace
Uses LESS & .479 & .274 & .074 & 1.749 & .081 & .005 & .005 [-.006, .019] \\ \addlinespace
Uses Scrum of Scrums & .120 & .111 & .049 & 1.083 & .279 & .002 & .002 [-.005, .011] \\ \addlinespace
Custom approach & .123 & .108 & .051 & 1.132 & .258 & .002 & .002 [-.005, .011] \\ \addlinespace
Other approach** & .424 & .140 & .133 & 3.025 & .003 & .016 & .016 [-.004, .040] \\ \addlinespace
Control: Team Experience with Agile** & .253 & .038 & .284 & 6.632 & .000 & .076 & .082 [.041, .130] \\ \addlinespace
Control: Size of organization & -.019 & .044 & -.019 & -.431 & .667 & .000 & .000 [-.003, .004] \\
\bottomrule
\end{tabular}
\vspace{2em}
\end{table}

\subsubsection{Does the scaling approach influence stakeholder satisfaction with responsiveness?}
We performed regression analysis to predict how satisfied stakeholders are with the responsiveness of teams based on the scaling approach in use (see Table~\ref{tab:stakeholdersatisfactionwithresponsivenessbyscalingapproach}). The regression model was significant, $F(7,522) = 5.238, p < .01$, and explained 6.4\% of the observed variance in responsiveness compared to 1.9\% by a secondary regression without control variables. The overall effect size of the model is qualified as \textit{small}, $f^2 = .064, 90\% CI [.030, .113]$. 

None of the scaling approaches significantly predict the satisfaction of responsiveness by stakeholders. This means that stakeholder satisfaction in this area can be high or low for teams, regardless of the chosen scaling approach.

Of the control variables, only the experience of teams with Agile has a significant effect that is qualified as \textit{small} based on its size, $beta = .213, p < .01, f^2 = .044, 90\% CI [.013, .082]$. So experienced teams are a little more able to satisfy their stakeholders in this area, although this effect is small and determined more by other factors.

\begin{table}[!ht]
\centering
\small
\caption{Coefficients, Standard Errors (SE), Beta's, t-values, significance, explained variance ($R^2$) and effect size ($f^2$) with 90\% confidence intervals for the chosen scaling approach, experience of teams with Agile, and the size of the organization on stakeholder satisfaction with responsiveness. *: statistically significant at $p <.05$. **: statistically significant at $p <.01$}
\label{tab:stakeholdersatisfactionwithresponsivenessbyscalingapproach}
\begin{tabular}{@{}m{3cm}m{1cm}m{1cm}m{1cm}m{1cm}m{1cm}m{1cm}m{2.5cm}@{}}
\toprule
\textbf{Variable} & \textbf{Coeff.} & \textbf{SE} & \textbf{$\beta$} & \textbf{t} & \textbf{$p$} & \textbf{$R^2$} & \textbf{$f^2$ / 90\% CI} \\
\midrule
Intercept** & 4.318 & .217 & .000 & 19.859 & .000 &  &  \\ \addlinespace
Uses SAFe & .027 & .142 & .009 & .193 & .847 & .000 & .000 [-.001, .002] \\ \addlinespace
Uses LESS & .220 & .276 & .034 & .795 & .427 & .001 & .001 [-.004, .008] \\ \addlinespace
Uses Scrum of Scrums & .201 & .112 & .084 & 1.795 & .073 & .006 & .006 [-.006, .020] \\ \addlinespace
Custom approach & -.031 & .109 & -.013 & -.286 & .775 & .000 & .000 [-.002, .002] \\ \addlinespace
Other approach & .190 & .142 & .060 & 1.344 & .180 & .003 & .003 [-.006, .014] \\ \addlinespace
Control: Team Experience with Agile** & .187 & .038 & .213 & 4.868 & .000 & .042 & .044 [.013, .082] \\ \addlinespace
Control: Size of organization & .049 & .044 & .050 & 1.117 & .264 & .002 & .002 [-.005, .011] \\
\bottomrule
\end{tabular}
\vspace{2em}
\end{table}

\subsubsection{Does the scaling approach influence stakeholder satisfaction with release frequency?}
We performed regression analysis to predict how satisfied stakeholders are with the release frequency of teams based on the scaling approach in use (see Table~\ref{tab:stakeholdersatisfactionwithreleasefrequencybyscalingapproach}). The regression model was significant, $F(7,521) = 6.531, p < .01$, and explained 8.1\% of the observed variance in responsiveness compared to 0.9\% by a secondary regression without control variables. The overall effect size of the model is qualified as \textit{small}, $f^2 = .088, 90\% CI [.046, .130]$. 

None of the scaling approaches significantly predict the satisfaction of release frequency by stakeholders. This means that stakeholder satisfaction in this area can be high or low for teams, regardless of the chosen scaling approach.

Of the control variables, only the experience of teams with Agile has a significant effect that is qualified as \textit{small} based on its size, $beta = .275, p < .01, f^2 = .077, 90\% CI [.037, .116]$. So experienced teams are a little more able to satisfy their stakeholders in this area, although this effect is small and determined more by other factors.

\begin{table}[!ht]
\centering
\small
\caption{Coefficients, Standard Errors (SE), Beta's, t-values, significance, explained variance ($R^2$) and effect size ($f^2$) with 90\% confidence intervals for the chosen scaling approach, experience of teams with Agile, and the size of the organization on stakeholder satisfaction with release frequency. *: statistically significant at $p <.05$. **: statistically significant at $p <.01$}
\label{tab:stakeholdersatisfactionwithreleasefrequencybyscalingapproach}
\begin{tabular}{@{}m{3cm}m{1cm}m{1cm}m{1cm}m{1cm}m{1cm}m{1cm}m{2.5cm}@{}}
\toprule
\textbf{Variable} & \textbf{Coeff.} & \textbf{SE} & \textbf{$\beta$} & \textbf{t} & \textbf{$p$} & \textbf{$R^2$} & \textbf{$f^2$ / 90\% CI} \\
\midrule
Intercept** & 3.814 & .258 & .000 & 14.774 & .000 &  &  \\ \addlinespace
Uses SAFe & -.074 & .168 & -.020 & -.438 & .661 & .000 & .000 [-.003, .003] \\ \addlinespace
Uses LESS & .488 & .328 & .063 & 1.486 & .138 & .004 & .004 [-.006, .014] \\ \addlinespace
Uses Scrum of Scrums & .024 & .133 & .008 & .178 & .859 & .000 & .000 [-.001, .001] \\ \addlinespace
Custom approach & .051 & .130 & .018 & .395 & .693 & .000 & .000 [-.002, .003] \\ \addlinespace
Other approach & .225 & .168 & .060 & 1.340 & .181 & .003 & .003 [-.006, .012] \\ \addlinespace
Control: Team Experience with Agile** & .290 & .046 & .275 & 6.347 & .000 & .071 & .077 [.037, .116] \\ \addlinespace
Control: Size of organization & -.047 & .052 & -.039 & -.888 & .375 & .001 & .001 [-.005, .007] \\
\bottomrule
\end{tabular}
\vspace{2em}
\end{table}

\subsubsection{Summary for stakeholder satisfaction}
A summary of the betas (i.e., indicating the direction and strength of the relationships) and effect sizes ($f^2$) that were identified in the regression analyses in this section is provided in Table~\ref{tab:stakeholdersatisfactionsummary} along with their significance and effect size classifications~\cite{cohen2013statistical}. After controlling for team experience and organization size, we found \textbf{one significant positive effect of moderate size from ``Other approach'' on the satisfaction of stakeholders with value. Another small effect was found for team experience on all indicators}. This suggests that stakeholder satisfaction is generally determined by factors other than the ones we investigated. While the scaling approach has some influence, it is very small and probably not practically relevant.

\begin{table}[!ht]
\centering
\small
\caption{Summary of beta's and effect sizes ($f^2$) by dimensions of stakeholder satisfaction for scaling approaches and control variables along with significance and effect size classification. *: statistically significant at $p <.05$. **: statistically significant at $p <.01$. a: no effect ($f^2 < .02$), b: small effect ($f^2 >= .02$), c: moderate effect ($f^2 >= .15$), d: large effect ($f^2 >= .35$)}
\label{tab:stakeholdersatisfactionsummary}
\begin{tabular}{@{}m{2.5cm}m{1cm}m{1cm}m{1cm}m{1cm}m{1cm}m{1cm}@{}}
\toprule
\textbf{Variable} & \textbf{Value} & & \multicolumn{2}{c}{\textbf{Responsiveness}} & \multicolumn{2}{c}{\textbf{Release Frequency}} \\
\cmidrule(lr){4-5} \cmidrule(lr){6-7}
 & $\beta$ & $f^2$ & $\beta$ & $f^2$ & $\beta$ & $f^2$ \\
\midrule
SAFe & .011 & $.000^a$ & .009 & $.000^a$ & -.020 & $.000^a$ \\ \addlinespace
LeSS & .074 & $.005^a$ & .034 & $.001^a$ & .063 & $.004^a$ \\ \addlinespace
Scrum of Scrums & .049 & $.002^a$ & .084 & $.006^a$ & .008 & $.000^a$ \\ \addlinespace
Custom approach & .051 & $.002^a$ & -.013 & $.000^a$ & .018 & $.000^a$ \\ \addlinespace
Other approach & .133** & $.016^a$ & .060 & $.003^a$ & .060 & $.003^a$ \\ \addlinespace
Control: Team Experience with Agile & .284** & $.082^b$ & .213** & $.044^b$ & .275** & $.077^b$ \\ \addlinespace
Control: Size of organization & -.019 & $.000^a$ & .050 & $.002^a$ & -.039 & $.001^a$ \\
\bottomrule
\end{tabular}
\vspace{2em}
\end{table}

\section{Discussion}
\label{sec:Discussion}
This study investigated to what extent the effectiveness of Agile teams is influenced by the Agile scaling approach in use. We begin with a summary of our results, first for team effectiveness and its core processes as reported by 12,534 team members from 4,013 teams, and then for the satisfaction with the outcomes of 529 of these teams as reported by 1,841 stakeholders.

\subsection{The influence of scaling approach on team effectiveness}
First, for team effectiveness and the core processes that determine it~\cite{verwijs2023theory}, we found small but significant differences between the scaling approaches for responsiveness, stakeholder concern, continuous improvement, team autonomy, management support, and overall team effectiveness. However, those effects mostly disappeared in the regression analyses that controlled for the experience of teams and organization size. While some significant effects were still observed, their standardized beta coefficients were very small and ranged between -.070 and .060. Moreover, their effect size ($f^2$) was so small as to be qualified as ``none''. For example, the difference between a team that uses ``SAFe'' as their scaling approach and a team that does not use scaling is only -.042 for responsiveness, or -.020 for overall team effectiveness. Similarly, teams that use ``Scrum of Scrums'' as their scaling approach compared to teams that do not use scaling see an increase of .057 in stakeholder concern. While such differences are statistically significant, their effect is too small to be practically relevant. We conclude from these results that \textbf{there is no meaningful effect of scaling approaches on the core processes of Agile team effectiveness} (responsiveness, stakeholder concern, continuous improvement, team autonomy, management support, and overall effectiveness).

\subsection{The influence of scaling approach on stakeholder satisfaction}
With respect to the satisfaction reported by stakeholders with team outcomes, a small but significant mean difference was found for delivered value, but not for responsiveness and release frequency. However, those effects mostly disappeared in the regression analyses that controlled for the experience of teams and organization size. The only exception was one significant effect we observed on the satisfaction of stakeholders with value delivered by teams that use a scaling approach other than the ones we analyzed specifically (``Other approach''). This standardized beta coefficient for this effect was .151 and its size was qualified as ``moderate''. We conclude from these results that, overall, \textbf{there is no meaningful effect of scaling approaches on the satisfaction of stakeholders with team outcomes} (on value, responsiveness, and release frequency).

\subsection{The influence of experience and organization size (control variables)}
Of our control variables, the experience that teams have with Agile contributed most strongly with effect sizes ranging between moderate and large. Organization size showed some significant effects, but all were qualified as \textit{small} based on their effect size. Finally, we observed that regression models that only considered the scaling approach tended to explain about 1-2\% of the variance, whereas regression models that included the control variables accounted up 47.5\% of the variance. This further underscores that scaling approaches alone explain very little variance in team effectiveness and stakeholder satisfaction and that context variables (such as experience of teams with Agile) explain much more variance. Thus we conclude that \textbf{teams appear to be comparably effective under any scaling approach and their stakeholders are equally satisfied}. Instead, the variance observed is attributable to other factors, one of which is the experiences of teams with Agile. Such team-level factors are clearly more relevant.

\subsection{Why Agile scaling approaches don't seem to make a difference}
This study confirms the pattern that emerged from qualitative comparisons of scaling approaches based on case studies, namely that no approach is clearly superior. Moreover, lightweight approaches do not seem to outperform more expansive and predictive approaches, which is a prevalent belief among practitioners~\cite{wolpers2023safe,safedelusion}. We will now explore potential explanations for our findings.

First, it is possible that the correlation between what is expected by the various approaches and what actually goes on in and around teams is low. All scaling approaches presume to encourage behavior that is consistent with the Agile manifesto~\cite{AgileManifesto2001}, such as frequent releases and close collaboration with stakeholders. One assumption behind the adoption of any scaling approach is that it will lead to such behavior. One example of this is the extent to which teams collaborate with stakeholders like users and customers. This involves such behaviors as asking clarifying questions to stakeholders, inviting them to provide feedback, and spending time with them to learn what is needed. Our measure for stakeholder concern specifically assessed the presence of such behaviors, with questions such as ``People in this team closely collaborate with users, customers, and other stakeholders.'' and ``The Product Owner of this team uses the Sprint Review to collect feedback from stakeholders''. However, our results show that the presence of such behaviors varies within scaling approaches, but not substantially between scaling approaches. Furthermore, the authors have observed teams in settings with ``LeSS'', ``SAFe'', ``Scrum of Scrums'', and even unscaled Scrum that were either not allowed to interact with stakeholders directly or applied the term ``stakeholder'' to internal roles, like a project manager or product owner, that did not interact with users or customers directly either. 

Another example of this is responsiveness. Frequent delivery is indeed a critical success factor of Agile projects~\cite{chow2008survey,verwijs2023theory}. Our measure for responsiveness assessed this with questions like ``The majority of the Sprints of this team result in an increment that can be delivered to stakeholders.'' and ``For this team, most of the Sprints result in an increment that can be released to users.''. While substantial variation was observed within scaling approaches, there was no substantial difference between scaling approaches. This may be an issue with language. Organizations may vary in what they define as a ``delivery'' and what is ``frequent''. The authors have informally observed cases where a release to a staging environment was labeled as a ``release'', even though internal procedures and processes restricted the actual release to a production environment to once per year. The actual behavior here is not what is intended by the Agile manifesto~\cite{AgileManifesto2001}, which aims to validate assumptions through frequent releases to users.

Taken together, this challenges the assumption that scaling approaches themselves change behavior in and around teams to be more consistent with the Agile manifesto. As the examples illustrate, each approach can be implemented in a manner that only results in superficial change, with different labels and different roles, but no deeper change of behavior to be more consistent with Agile development practices. Other factors seem more important in encouraging that behavior, such as experience with Agile.

A second explanation may lie in adherence. Organizations and teams may vary in the degree to which they follow what is prescribed in their scaling approach of choice. Larger differences may be observed when the analyses control for the degree to which all parts of the approach are practiced, and not a selection of them. However, approaches like ``SAFe'' and ``LeSS'' specifically state that they are modular and that organizations should select the elements that work for them~\cite{SAFewebsite,LESSWebsite}. Another approach would be to assess adherence to the principles of the Agile manifesto~\cite{AgileManifesto2001}. However this is conceptually close to the core processes of team effectiveness we used in this study, and we did not find meaningful differences between scaling approaches. Moreover, this would reinforce that the approach itself does not make the difference, but the degree to which organizations honor the principles of Agile software development (i.e. collaborate closely with users, and release to them frequently) which may be mostly orthogonal to the scaling approach.

Finally, it is possible that the number of teams moderates the association between team effectiveness and stakeholder satisfaction on the one hand and the scaling approach on the other. As the number of teams grows and coordination becomes more complicated, this may put more strain on the scaling approach~\cite{Ebert17} Whereas a complex approach like ``SAFe'' may provide more structure and guidance to take this strain, simpler approaches like ``LeSS'' and ``Scrum of Scrums'' may provide less support. This seems particularly relevant for organizations with limited experience with Agile. Future investigations can control for both experience and the number of teams to account for test alternative explanations.

\subsection{Implications for practice}
We now turn to the implications of our study and what they mean in the day-to-day practice of professionals and organizations that attempt to scale Agile methodologies.

The first implication is that \textbf{there is no such ``best'' scaling approach}, especially if concerned with team effectiveness and stakeholder satisfaction. Any variation in these variables is attributable to other factors. This also means that teams can be similarly effective under any scaling approach, and have equally satisfied stakeholders. Thus, we recommend that practitioners prioritize those factors that have been empirically linked to team effectiveness and stakeholder satisfaction and worry less about which scaling approach to pick. This includes factors such as continuous improvement~\cite{verwijs2023theory}, psychological safety~\cite{edmondson2014psychological}, inter-team collaboration~\cite{riedel2021,dingsoyr2013research}, teamwork~\cite{strode2022teamwork}, team autonomy~\cite{junker2021Agile,verwijs2023theory} and socio-technical skills of developers~\cite{verwijs2023theory}. Verwijs \& Russo~\cite{verwijs2023theory} were able to explain up to 75.6\% of the variance of team effectiveness with team autonomy, a climate of continuous improvement, concern for stakeholders, responsiveness, and management support.

Second, we follow the conclusions of Almeida \& Espinheira~\cite{Almeida21} and recommend that practitioners \textbf{pick the scaling approach that best suits the culture, structure, and experience of their organization}. The comprehensiveness of ``SAFe'' may work better in highly regulated, corporate settings that have limited experience with Agile development, whereas the simplicity of ``Scrum of Scrums'' or ``LeSS'' may be more suited to organizations that are already familiar with it. Moreover, complex approaches to scaling like ``SAFe'' and ``Disciplined Agile'' provide more guidance for governance, release planning, budgeting, portfolio planning, and technical practices, whereas simpler approaches like ``LeSS'' and certainly ``Scrum of Scrums'' leave this open. To illustrate this, Ciancarini et. al.~\cite{Ciancarini22} conclude from a multivocal literature review and a survey among practitioners that the comprehensiveness and informed support of organizations by ``SAFe'' is the primary reason for its success. Organizations have to be cognizant of the gap between their current state and the desired state of Agility in such areas. If this gap is too large, a comprehensive approach may help organizations ease into it, whereas a simpler approach may leave so much open that it creates more confusion than clarity. Once organizations build experience with Agile development methodologies, they can transition into simpler approaches or develop their own. Thus, we propose that organizations select for goodness-of-fit instead of simplicity alone and periodically reflect on the extent to which their scaling approach allows or impedes teams to practice the principles of Agile (software) development~\cite{AgileManifesto2001}

Third, we recommend that practitioners \textbf{monitor stakeholder satisfaction and team effectiveness regardless of their scaling approach}. Our results show substantial variation in these areas within each scaling approach, though not between. We also recommend monitoring the extent to which the behaviors observed in and around teams reflect the principles of the Agile manifesto~\cite{AgileManifesto2001}. This includes behaviors around stakeholder collaboration, collaborative goal-setting, frequent releases to production, expanding team autonomy, and continuously reflecting and improving the process by which teams deliver to stakeholders.

Fourth, our results do not support the anecdotal negative opinion of complex approaches like ``SAFe'' we observed among Agile practitioners~\cite{wolpers2023safe,safedelusion}. It is likely that practitioners use different criteria, such as simplicity, personal preferences, or favor approaches with lower prescriptiveness. However, Ciancarini et. al.~\cite{Ciancarini22} found that practitioners of ``SAFe'' do not consistently experience it as too complex, too rigid, inhibiting learning and improvement, or too hierarchical. Another possibility is that practitioners have a broader comparative experience with different approaches, whereas the subjects in our study - team members and stakeholders - generally have experience only with the approach in use in their organization. We can not rule out that stakeholders of teams that use ``SAFe'' would be more satisfied with the value delivered by teams, their responsiveness, and release frequency under a simpler approach like  ``LeSS'' or ``Scrum of Scrums''. Unfortunately, such a hypothesis is hard to test as few stakeholders are in a position to experience one scaling approach with a team and then another consecutively. We also note that we did not observe meaningful differences in the core processes of team effectiveness between scaling approaches. Since these indicators have been found to explain a substantial amount of the variance in the effectiveness of Agile teams~\cite{verwijs2023theory}, we believe it is more likely to expect comparable results.

Finally, the role of experiences with Agile emerged as a surprisingly strong predictor of team effectiveness and a moderate one for stakeholder satisfaction in this study. In contrast to the scaling approach, this factor does meaningfully and positively impact the extent to which teams engage with stakeholders, are responsive, engage in continuous improvement, capitalize on their autonomy, and experience more support from management. Experienced teams also tend to have more satisfied stakeholders. This may be closely related to what is called an ``Agile mindset'' (AM) by Eilers, Peters \& Leimeister~\cite{eilers2022agile}. They define it as consisting of an attitude towards learning, collaborative exchange, empowered self-guidance, and customer co-creation. Indeed, experienced teams in our study also show higher responsiveness, stakeholder concern, team autonomy, continuous improvement, management support, and overall team effectiveness. The presence of such a mindset in and around teams may be much more relevant to team and business outcomes, regardless of the scaling approach, as it allows teams to better deal with volatility, uncertainty, complexity, and ambiguity (VUCA)~\cite{eilers2022agile}. The notion of an AM also provides a more fine-grained set of variables to investigate compared to the course-grained measure for Agile experience we used in this study. Thus, future studies can attempt to replicate our results with AM as a control variable. Finally, we can not conclusively establish causality from a cross-sectional study such as this one, but it does suggest that \textbf{broadening that experience with Agile through training, coaching, and practice} is an effective recommendation.

A summary of our core findings and implications are provided in Table~\ref{tab:Findings}.

\begin{table}[!ht]
\centering
\small
\caption{Summary of key findings \& implications}
\label{tab:Findings}
\begin{tabular}{@{}m{2.5cm}m{4.5cm}m{4.5cm}@{}}
\toprule
    & \textbf{Findings} & \textbf{Implications} \\
\midrule
\textbf{Comparison of team effectiveness by scaling approach} & 4,013 Agile teams were compared by scaling approach (``SAFe'', ``LeSS'', ``Scrum of Scrums'', ``Custom approach'', other approaches, and no scaling) on team effectiveness and five core processes that contribute to it; responsiveness, stakeholder concern, continuous improvement, team autonomy, and management support. Small and significant differences were observed in all variables. However, the effects mostly disappeared when we controlled for the experience of teams and organization size. While some significant effects were still observed, their effect size ($f^2$) was so small as to be qualified as ``no effect'' ($f^2 < .02$). & Scaling approach does not appear to be a meaningful predictor of team effectiveness or the processes that shape it. Teams can be comparably effective in ``SAFe'', ``LeSS'', or the other approaches studied. We recommend that practitioners pick the approach that suits their needs and monitor and expand their Agility from there. \\ \addlinespace
\textbf{Comparison of stakeholder satisfaction by scaling approach} & 529 Agile teams were compared by scaling approach on how satisfied their stakeholders were with delivered value, responsiveness, and release frequency. A small and significant difference was observed for value, but not for responsiveness or release frequency. As with team effectiveness, the effects mostly disappeared when we controlled for the experience of teams and organization size. While some significant effects were still observed, their effect size ($f^2$) was so small as to be qualified as ``no effect'' ($f^2 < .02$). & Scaling approach does not appear to be a meaningful predictor of stakeholder satisfaction. Stakeholders can be comparably satisfied in ``SAFe'', ``LeSS'', or the other approaches studied. We recommend that practitioners pick the approach that suits their needs and monitor stakeholder satisfaction to identify improvement in factors that do contribute.\\ \addlinespace
\textbf{Role of experience and organization size} & We observed that regression models that only considered the scaling approach tended to explain about 1-2\% of the variance, whereas regression models that included the control variables accounted up 47.5\% of the variance. This was mostly due to team experience, with standardized coefficients ranging between .213 and .686. & Regardless of the scaling approach chosen, teams that build experience with Agile tend to be more effective and have more satisfied stakeholders. Broadening that experience through training, coaching, and exposure is a useful recommendation. \\ \addlinespace
\bottomrule
\end{tabular}
\vspace{2em}
\end{table}

\subsection{Limitations}
\label{sec:limitations}
In this section, we discuss the threats to the validity of our sample study.
We published team-level data and syntax files to Zenodo for reproducibility~\footnote{The complete replication package is openly available under CC-BY-NC-SA 4.0 license on Zenodo, DOI: \url{https://doi.org/10.5281/zenodo.8396487}.}. 

\textbf{Internal validity}
Internal validity refers to the confidence with which changes in the dependent variables can be attributed to the independent variables and not other uncontrolled factors~\cite{cook1979quasi}. Several strategies were used to maximize internal validity. First, online questionnaires are prone to bias and self-selection as a result of their voluntary (non-probabilistic) nature. This was counteracted by embedding our questions in a tool that is regularly used by Agile software teams to self-diagnose their process and identify improvements. Team members were invited by people in their organization to participate. Teams invited their own stakeholders. Second, we thoroughly cleaned the dataset of careless responses to prevent them from influencing the results. Third, we did not inform the participants of our specific research questions to prevent them from answering in a socially desirable manner.

Despite our safeguards, there may still be confounding variables that we were unable to control for. This is particularly relevant to the operationalization of team effectiveness, which is based on self-reported scores on team morale and the perceived satisfaction of stakeholders. Mathieu et al. ~\cite{mathieu2008team} recognize that such affect-based measures may suffer from a ``halo effect''. We addressed this issue by also performing a secondary analysis that relied on the satisfaction as reported by stakeholders themselves for those teams where such evaluations were available in the tool. A moderate correlation was found between the satisfaction of stakeholders as reported by team members and the satisfaction reported by stakeholders directly (between .346 and .424). While this provides some evidence of a halo effect, the measure used with stakeholders was more extensive and multi-dimensional whereas the measure used with team members only asked to what extent they believed their stakeholders to be satisfied.

Several confounding factors have been identified that we could not control for. The first is that there may be a selection bias in which stakeholders are invited. We cannot conclusively rule out that teams only invited stakeholders that they assumed would be satisfied and ignored those who would not be. Similarly, it is possible that only highly effective teams invited their stakeholders whereas less effective teams did not. However, a post-hoc test did not reveal a significant effect of team effectiveness on the number of stakeholders invited, $F(1,422) = .155, p = .69$.

\textbf{Construct validity}
Construct validity refers to the degree to which the measures used in a study measure their intended constructs~\cite{cook1979quasi}. To measure the indicators of team effectiveness, we relied on an existing questionnaire that was developed and tested earlier in Verwijs \& Russo~\cite{verwijs2023theory}. The questionnaire to evaluate the satisfaction of stakeholders was developed for this and future investigations. 

A confirmatory factor analysis (CFA) showed that all items were loaded primarily on their intended scales (see Table 3 in the Appendix). A heterotrait-monotrait analysis (HTMT) yielded no issues. This means that our measures are distinguishable from each other and that any overlap does not confound the results. The reliability for all measures exceeded the cutoff recommended in the literature ($CR>=.70$~\cite{hair2019multivariate}), except social desirability. Thus, we are confident that we reliably measured the intended constructs.

\textbf{Conclusion validity}
Conclusion validity assesses the extent to which the conclusions about the relationships between variables are reasonable based on the results~\cite{cozby2012methods}. Our sample was also large enough to identify medium effects ($f=.15$) with a statistical power of 96\%. For the comparison of stakeholder satisfaction between scaling approaches, we do note that the group for LeSS contained only 7 teams, representing 44 stakeholders.

\textbf{External validity}
Finally, external validity concerns the extent to which the results actually represent the broader population~\cite{goodwin2016research}. First, we assess the ecological validity of our results to be high. Our questionnaire was integrated into a more general tool that Agile software teams use to improve their processes. Participants were invited by people in their organization, usually Scrum Masters. Thus, the data is more likely to reflect realistic teams than a stand-alone questionnaire or an experimental design. 

We do not know how well our sample reflects the total population. However, our sample composition (Table~\ref{tab:samplecomposition}) shows that a wide range of teams participated in the questionnaire, with different levels of experience from different parts of the world and different types of organizations. We also observed a broad range of scores on the various measures. This provides confidence that a wide range of teams participated. Furthermore, our sample size and the aggregation of individual-level responses to team-level aggregates reduce variability due to non-systematic individual bias.

\subsection{Future research}
This study focused on team-level effectiveness and team-level stakeholder satisfaction. Although this paper contributes to the understanding of organization-level outcomes, it would be meaningful for future studies to investigate how organizational outcomes vary by scaling approach (e.g., financial baseline, market share, revenue). Such investigations would contribute to a more comprehensive picture of how various scaling approaches affect organizations.

Future research can also investigate what happens when teams or organizations switch from one approach to another. How does such a change affect team-level effectiveness, stakeholder satisfaction, and organizational outcomes? That kind of longitudinal data would allow researchers to determine whether the level of reported satisfaction of stakeholders would be different if they had prior experience with other approaches.

Future research can offset the costs and benefits of the various scaling approaches. If the choice of scaling approach does not correlate with actual team effectiveness or stakeholder satisfaction to a meaningful degree, it would be economical to pick the option with the lowest implementation costs. This is one area where the approaches vary substantially. ``SAFe'' requires additional training and certification, along with organizational changes, whereas ``Scrum of Scrums'' requires neither.

Finally, we recognize in our discussion that organizations probably do well in selecting a scaling approach based on contingency factors instead of simplicity alone. Simple and lightweight approaches like ``LeSS'', ``Nexus'' and particularly ``Scrum of Scrums'' may leave too much open for organizations with very little experience with Agile, leading to confusion and uncertainty. A more prescriptive and comprehensive approach like ``SAFe'', ``Disciplined Agile'' may offer more guidance here. Future investigations can develop evidence-based models to help organizations determine their goodness-of-fit with various scaling approaches. For example, this could include factors such as organizational culture, prior experience with Agile, leadership styles, budget structure, planning cycles, and regulatory requirements. One such model is proposed by Laanti~\cite{laanti2017agile}. This model identifies five progressive levels of successful scaling of Agile methodologies. Each level addresses how work is scaled and what benefits organizations gain from scaling. For example, organizations on the first level have the basics in place, such as Product Backlog tool, a prioritized Backlog and apply a framework like Scrum. On the other hand, organizations at the highest level have developed their own approach to scaling and release new increments on a daily or even hourly basis. This agility is leveraged to expand into new markets, build new businesses and outperform competitors. 

\section{Conclusion}
\label{sec:Conclusion}
Agile scaling approaches have become increasingly popular as (software) projects become more complex~\cite{mishra2011}. Such approaches have been developed to address a perceived gap in Agile methodologies; namely how to scale Agile development from one team to many teams. Of these approaches, the Scaled Agile Framework (``SAFe'') is the most popular~\cite{SAFe2018litrev,Conboy19} although it is also seen as the most complex one~\cite{Ebert17}. Other well-known scaling approaches are Large Scale Scrum (``LeSS'') and ``Scrum of Scrums''~\cite{schwaber2004agile}. But many organizations also develop their own Agile scaling approach. There is some anecdotal evidence that practitioners prefer simpler approaches over more complex ones such as SAFe~\cite{wolpers2023safe,safedelusion}. 

Several studies have investigated the success factors and risks of the various scaling approaches, mostly based on qualitative methods e.g., interviews with practitioners or case studies. Each scaling approach has its own challenges, but no approach appears to be consistently better~\cite{Almeida21,Edison22,Putta18}. To date, no studies have systematically compared Agile scaling approaches based on empirical data from a consistent measure. The aim of this study was to investigate if certain Agile scaling approaches are more effective than others. We conducted a survey study among 11,376 team members grouped into 4,013 Agile teams to assess their effectiveness and the five core processes that give rise to it. Furthermore, stakeholder satisfaction was reported by 1,841 stakeholders for 529 of these teams.

While our results yielded some statistically significant differences, both the absolute differences and their effect size were small to non-existent. This applied both to the five indicators of team effectiveness as well as four dimensions of stakeholder satisfaction as reported by stakeholders themselves. We found that any observed differences often diminished when we controlled for the experience of teams with Agile and, to a lesser extent, the size of organizations. Thus, we conclude that the scaling approach itself is not a meaningful predictor of team effectiveness and stakeholder satisfaction in a practical sense. Teams that use ``SAFe'' appear to be equally capable to be effective and satisfy stakeholders than teams that use ``LeSS'', a custom approach, ``Scrum of Scrums'' or another scaling approach.

Our findings are consistent with prior investigations by Almeida \& Espinheira~\cite{Almeida21} and Edison, Wang \& Conboy~\cite{Edison22}. Without strong evidence that shows clear differences between scaling approaches, we feel that the evidence-based recommendation is for organizations to use the scaling approach that works for them and does not create too much of a mismatch between mindset, structure, and processes. Stakeholder satisfaction and team effectiveness can then be monitored to identify areas for improvement and provide training and coaching to expand the experience of teams with Agile methodologies.

\section*{Supplementary Materials} 
The complete replication package is openly available under CC-BY-NC-SA 4.0 license on Zenodo, DOI: \url{https://doi.org/10.5281/zenodo.8396487}.


\section*{Acknowledgment}
The authors would like to thank Rasmus Broholm for feedback on an earlier version of this manuscript.

\section*{Conflict of Interest}
Christiaan Verwijs declares a financial interest in The Liberatators BV.

\bibliographystyle{spmpsci}   
\bibliography{main}  

\clearpage 
\appendix

\section{Appendix: Questionnaire}
\label{sec:appendix:questionnaire}

\begin{longtable}{@{}m{4cm}m{4cm}m{4cm}@{}}
\caption{Measurement Instrument for team members.} \label{tab:measurementinstrument} \\
\toprule
Question & Label & Categories \\
\midrule
\endfirsthead

\multicolumn{3}{c}%
{{\bfseries \tablename\ \thetable{} -- continued from previous page}} \\
\toprule
Question & Label & Categories \\
\midrule
\endhead

\bottomrule
\multicolumn{3}{r}{{Continued on next page}} \\
\endfoot

\bottomrule
\endlastfoot
TeamKey & Unique key for each participating team (automatically entered) & Hidden \\
\midrule
\multicolumn{3}{l}{\textbf{Demographic Variables}} \\
Location & Where is this team based? & Western Europe, Eastern Europe, North America, Central \& South America, Middle East, Oceania, Africa, South Asia, East Asia, South-East Asia, Global, Other \\
Organisation Sector & Which sector is this organisation mostly active in? & Technology and telecommunications, Professional and Business services,   Government, Manufacturing, Real estate,Retail, Healthcare, Non-profit,   Automotive, Financial, Other \\
Role & What role best describes what you do in this team? (why we ask this) & Scrum Master, Product Owner, Developer, Visual/UX  Designer, Tester, Marketeer or sales,   Analyst, Infrastructure engineer, Other, I do not want to say \\
Scaling Framework & When applicable, what approach does your organization primarily use to   coordinate work between multiple Scrum/Agile teams? & None, Scrum of Scrums, Scale Agile Framework (SAFe),   Large-Scale Scrum (LESS), Other approach, We created our own approach \\
Type Of Product & Where are the people that this team works for - like users and customers   - mostly based? & Mostly within our own organisation (e.g. internal use), Mostly outside   our own organisation (e.g. consumers, external customers or users) \\
Team Size & How many members does this team typically have, including a Scrum Master   and Product Owner? & 1-4 members, 5-10 members, 11-16 members, More than 16 members \\
\midrule
\multicolumn{3}{l}{\textbf{Responsiveness}} \\
Refinement 1 & The Sprint Backlog of this team usually contains many small items. & Likert (7), Completely disagree to Completely agree \\
Refinement 2 & During the Sprint, this team spends time to clarify work for the next   couple of Sprints. & Likert (7), Completely disagree to Completely agree \\
Refinement 3 & During the Sprint, this team spends time breaking down work for coming   Sprints. & Likert (7), Completely disagree to Completely agree \\
Release Automation 1 & The process this team uses to deploy software to production is mostly   automated. & Likert (7), Completely disagree to Completely agree \\
Release Automation 2 & A release to production can generally be performed without manual steps. & Likert (7), Completely disagree to Completely agree \\
Release Frequency 1 & The majority of the Sprints of this team result in software that can be   deployed to production. & Likert (7), Completely disagree to Completely agree \\
Release Frequency 2 & For this team, most of the Sprints result in an increment that can be   released to users. & Likert (7), Completely disagree to Completely agree \\
Release Frequency 3 & The majority of the Sprints of this team result in an increment that can   be delivered to stakeholders. & Likert (7), Completely disagree to Completely agree \\
\midrule
\multicolumn{3}{l}{\textbf{Stakeholder Concern}} \\
Shared Goals 1 & This team generally has clear Sprint Goals. & Likert (7), Completely disagree to Completely agree \\
Shared Goals 2 & During Sprint Planning, this team formulates a clear goal for the Sprint. & Likert (7), Completely disagree to Completely agree \\
Stakeholder Collaboration 1 & Members of this team frequently meet with users or customers of what this   team creates. & Likert (7), Completely disagree to Completely agree \\
Stakeholder Collaboration 2 & People from this team often invite or visit people that use what this   team works on. & Likert (7), Completely disagree to Completely agree \\
Stakeholder Collaboration 3 & People in this team closely collaborate with users, customers and other   stakeholders. & Likert (7), Completely disagree to Completely agree \\
Stakeholder Collaboration 4 & This team frequently runs experiments or workshops to discover how people   (want to) use the product. & Likert (7), Completely disagree to Completely agree \\
Stakeholder Collaboration 5 & Before implementing a feature fully, this team often tests simple   versions with users first. & Likert (7), Completely disagree to Completely agree \\
Sprint Review Quality 1 & The Product Owner of this team uses the Sprint Review to collect feedback   from stakeholders. & Likert (7), Completely disagree to Completely agree \\
Sprint Review Quality 2 & During Sprint Reviews, stakeholders frequently try out what this team has   been working on during the Sprint. & Likert (7), Completely disagree to Completely agree \\
Sprint Review Quality 3 & Most Sprint Reviews result in useful changes to the Product Backlog of   this team. & Likert (7), Completely disagree to Completely agree \\
Value Focus 1 & The Product Owner of this team has a clear vision for the product. & Likert (7), Completely disagree to Completely agree \\
Value Focus 2 & The Product Backlog of this team is ordered with a strategy in mind. & Likert (7), Completely disagree to Completely agree \\
Value Focus 3 & Everyone in this team is familiar with the vision for the product. & Likert (7), Completely disagree to Completely agree \\
\midrule
\multicolumn{3}{l}{\textbf{Continuous Improvement}} \\
Learning Environment 1 & In and around this team, people are given time to support learning. & Likert (7), Completely disagree to Completely agree \\
Learning Environment 2 & In and around this team, people are rewarded for learning. & Likert (7), Completely disagree to Completely agree \\
Learning Environment 3 & In and around this team, people are encouraged to learn new skills,   techniques or practices. & Likert (7), Completely disagree to Completely agree \\
Learning Environment 4 & In and around this team, people see learning as a part of their work. & Likert (7), Completely disagree to Completely agree \\
Psychological Safety 1 & In and around this team, people give open and honest feedback to each   other. & Likert (7), Completely disagree to Completely agree \\
Psychological Safety 2 & In and around this team, people listen to the others' views before   speaking. & Likert (7), Completely disagree to Completely agree \\
Psychological Safety 3 & In and around this team, whenever people state their view, they also ask   what others think. & Likert (7), Completely disagree to Completely agree \\
Metric Usage 1 & This team often inspects metrics to identify process improvements. & Likert (7), Completely disagree to Completely agree \\
Metric Usage 2 & Decisions about what this team does are often influenced by metrics. & Likert (7), Completely disagree to Completely agree \\
Quality 1 & Members of this team have a shared understanding of what quality means to   them. & Likert (7), Completely disagree to Completely agree \\
Quality 2 & People in this team frequently talk about quality and how to improve it. & Likert (7), Completely disagree to Completely agree \\
Quality 3 & This team is always looking for ways to improve quality. & Likert (7), Completely disagree to Completely agree \\
Shared Learning 1 & This team frequently works with other groups or teams to solve shared   problems. & Likert (7), Completely disagree to Completely agree \\
Shared Learning 2 & Teams in this organization share what they learn with other teams. & Likert (7), Completely disagree to Completely agree \\
Shared Learning 3 & Members from this team frequently meet with other teams to identify   improvements. & Likert (7), Completely disagree to Completely agree \\
Sprint Retrospective Quality 1 & The Sprint Retrospectives of this team generally result in at least one   useful improvement. & Likert (7), Completely disagree to Completely agree \\
Sprint Retrospective Quality 2 & During Sprint Retrospectives, this team openly talks about improvements. & Likert (7), Completely disagree to Completely agree \\
Sprint Retrospective Quality 3 & This team uses Sprint Retrospectives to explore solutions for persistent   challenges. & Likert (7), Completely disagree to Completely agree \\
\midrule
\multicolumn{3}{l}{\textbf{Team Autonomy}} \\
Cross Functionality 1 & Most people in this team have the ability to solve the problems that come   up in their work. & Likert (7), Completely disagree to Completely agree \\
Cross Fucntionality 2 & Everyone in this team has more than enough training and experience for   the kind of work they have to do. & Likert (7), Completely disagree to Completely agree \\
Self Management 1 & This team has control over the scheduling of teamwork. & Likert (7), Completely disagree to Completely agree \\
Self Management 2 & This team is free to choose the method(s) to use in carrying out work. & Likert (7), Completely disagree to Completely agree \\
Self Management 3 & This team is able to choose the way to go about its work. & Likert (7), Completely disagree to Completely agree \\
\midrule
\multicolumn{3}{l}{\textbf{Management Support}} \\
Management Support 1 & People in a management position generally understand why this team works   with Scrum. & Likert (7), Completely disagree to Completely agree \\
Management Support 2 & People in a management position help this team work with Scrum. & Likert (7), Completely disagree to Completely agree \\
\midrule
\multicolumn{3}{l}{\textbf{Team Effectiveness}} \\
Stakeholder Satisfaction 1 & Stakeholders are generally happy with the software this team delivers. & Likert (7), Completely disagree to Completely agree \\
Stakeholder Satisfaction 2 & Stakeholders are generally happy with how fast this team responds to   their needs. & Likert (7), Completely disagree to Completely agree \\
Stakeholder Satisfaction 3 & Our stakeholders compliment us with the value that we deliver to them. & Likert (7), Completely disagree to Completely agree \\
Team Morale 1 & I am proud of the work that I do for this team. & Likert (7), Completely disagree to Completely agree \\
Team Morale 2 & I am enthusiastic about the work that I do for this team. & Likert (7), Completely disagree to Completely agree \\
Team Morale 3 & I find the work that I do for this team full of meaning and purpose. & Likert (7), Completely disagree to Completely agree \\
\midrule
\multicolumn{3}{l}{\textbf{Control Variables}} \\
Team Experience With Agile & I consider this team to be very experienced with Scrum / Agile. & Likert (7), Completely disagree to Completely agree \\
Organisation Size & What is the size of this organisation? & Between 1 and 50 employees, Between 51 and 500 employees, Between 501 and   5.000 employees, More than 5.000 employees, I have no idea \\
\end{longtable}

\begin{table}[!ht]
\centering
\small
\caption{Measurement Instrument for stakeholders.}
\label{tab:MeasurementInstrument}
\begin{tabular}{@{}m{4cm}m{4cm}m{4cm}@{}}
\toprule
\textbf{Question} & \textbf{Label} & \textbf{Categories} \\
\midrule
TeamKey & Unique key for each participating team (automatically entered) & Hidden \\ \addlinespace
\multicolumn{3}{l}{\underline{Demographic Variables}} \\ \addlinespace
Stakeholder Type & What best describes your stake in what this team delivers? & I (want to) use it in daily life, I pay for its development, I both (want to) use it in daily life and pay for its development, I am interested in it, but I do not use it nor pay for its development \\ \addlinespace
\multicolumn{3}{l}{\underline{Satisfaction Measures}} \\ \addlinespace
Satisfaction With Quality 1 & What this team delivers is of high quality. & Likert (7), Completely disagree to Completely agree \\ \addlinespace
Satisfaction With Quality 2 & I am satisfied with the quality of what this team delivers. & Likert (7), Completely disagree to Completely agree \\ \addlinespace
Satisfaction With Quality 3 & When the team delivers a new version, it is usually free of serious bugs. & Likert (7), Completely disagree to Completely agree \\ \addlinespace
Satisfaction With Responsiveness 1 & I frequently meet or interact with members of this team. & Likert (7), Completely disagree to Completely agree \\ \addlinespace
Satisfaction With Responsiveness 2 & I have a good sense of what this team is working on. & Likert (7), Completely disagree to Completely agree \\ \addlinespace
Satisfaction With Responsiveness 3 & When I have an idea or suggestion, members of the team are available to listen to me. & Likert (7), Completely disagree to Completely agree \\ \addlinespace
Satisfaction With Responsiveness 4 & The team frequently asks for my feedback, ideas or thoughts. & Likert (7), Completely disagree to Completely agree \\ \addlinespace
Satisfaction With Release Frequency 1 & This team frequently delivers new versions. & Likert (7), Completely disagree to Completely agree \\ \addlinespace
Satisfaction With Release Frequency 2 & I am satisfied with how often new versions are released. & Likert (7), Completely disagree to Completely agree \\ \addlinespace
Satisfaction With Release Frequency 3 & The frequency of new releases is good enough for my needs. & Likert (7), Completely disagree to Completely agree \\ \addlinespace
Satisfaction With Value 1 & I am satisfied with the value that this teams delivers. & Likert (7), Completely disagree to Completely agree \\ \addlinespace
Satisfaction With Value 2 & I am happy with the value that this team delivers every Sprint. & Likert (7), Completely disagree to Completely agree \\
\bottomrule
\end{tabular}
\vspace{2em}
\end{table}

\section{Appendix: Scale validation}
\label{sec:appendix:scalevalidation}

\begin{table}[!ht]
\centering
\small
\caption{Results of Confirmatory Factor Analysis from individual-level responses ($n=9,569$). Principal Components Analysis with Oblimin rotation and Kaiser normalization. Items and components are ordered by extraction. Factor loadings below .30 have been suppressed for readability.}
\label{tab:stakeholdercfa}
\begin{tabular}{@{}m{6cm}m{1.5cm}m{1.5cm}m{1.5cm}@{}}
\toprule
\textbf{Item} & \textbf{1} & \textbf{2} & \textbf{3} \\
\midrule
Satisfaction With Quality 2 & 0.876 &  &  \\ \addlinespace
Satisfaction With Quality 1 & 0.874 &  &  \\ \addlinespace
Satisfaction With Quality 3 & 0.860 &  &  \\ \addlinespace
Satisfaction With Value 1 & 0.657 &  &  \\ \addlinespace
Satisfaction With Value 2 & 0.558 &  & 0.329 \\ \addlinespace
Satisfaction With Responsiveness 1 &  & 0.905 &  \\ \addlinespace
Satisfaction With Responsiveness 4 &  & 0.817 &  \\ \addlinespace
Satisfaction With Responsiveness 3 &  & 0.714 &  \\ \addlinespace
Satisfaction With Responsiveness 2 &  & 0.655 &  \\ \addlinespace
Satisfaction With Release Frequency 3 &  &  & 0.907 \\ \addlinespace
Satisfaction With Release Frequency 2 &  &  & 0.864 \\
\bottomrule
\end{tabular}
\vspace{2em}
\end{table}

\newpage

\section{Appendix: Full Regression analyses}
\label{sec:appendix:regressionanalysis}

\begin{table}[!ht]
\centering
\small
\caption{Coefficients, Standard Errors (SE), Beta's, t-values, significance, explained variance ($R^2$) and effect size ($f^2$) with 90\% confidence intervals for the chosen scaling approach on the responsiveness of teams, first for an uncontrolled model and then for a model that controls for team experience with Agile and organization size. *: statistically significant at $p <.05$. **: statistically significant at $p <.01$}
\label{tab:regressionAnalysis}
\begin{tabular}{@{}m{3cm}m{1cm}m{1cm}m{1cm}m{1cm}m{1cm}m{1cm}m{2.5cm}@{}}
\toprule
\textbf{Variable} & \textbf{Estimate} & \textbf{SE} & \textbf{Beta} & \textbf{t} & \textbf{p} & \textbf{R2} & \textbf{f2 / 90\% CI} \\
\midrule
\underline{Regression model without control variables} &  &  &  &  &  & .009 & .009 [.003, .015] \\ \addlinespace
Intercept** & 4.419 & .029 & .000 & 154.802 & .000 &  &  \\ \addlinespace
Uses SAFe & .029 & .052 & .010 & .551 & .582 & .000 & .000 [.000, .001] \\ \addlinespace
Uses LESS & .090 & .093 & .016 & .965 & .335 & .000 & .000 [-.001, .001] \\ \addlinespace
Uses Scrum of Scrums** & .266 & .046 & .104 & 5.815 & .000 & .008 & .008 [.003, .014] \\ \addlinespace
Custom approach & .044 & .047 & .016 & .926 & .354 & .000 & .000 [-.001, .001] \\ \addlinespace
Other approach & .091 & .071 & .021 & 1.277 & .202 & .000 & .000 [-.001, .002] \\ \addlinespace
\underline{Regression model with control variables} &  &  &  &  &  & .352 & .543 [.519, .567] \\ \addlinespace
Intercept** & 2.799 & .058 & .000 & 48.297 & .000 &  &  \\ \addlinespace
Uses SAFe** & -.124 & .045 & -.042 & -2.777 & .006 & .001 & .001 [-.001, .004] \\ \addlinespace
Uses LESS & -.042 & .077 & -.007 & -.545 & .586 & .000 & .000 [.000, .000] \\ \addlinespace
Uses Scrum of Scrums & .003 & .039 & .001 & .066 & .947 & .000 & .000 [.000, .000] \\ \addlinespace
Custom approach* & -.084 & .040 & -.032 & -2.113 & .035 & .001 & .001 [-.001, .002] \\ \addlinespace
Other approach & -.045 & .060 & -.010 & -.751 & .453 & .000 & .000 [-.001, .001] \\ \addlinespace
Control: Team Experience with Agile** & .415 & .009 & .594 & 44.749 & .000 & .343 & .521 [.497, .545] \\ \addlinespace
Control: Size of organization** & -.044 & .015 & -.040 & -2.930 & .003 & .001 & .001 [-.001, .004] \\
\bottomrule
\end{tabular}
\vspace{2em}
\end{table}

\begin{table}[!ht]
\centering
\small
\caption{Coefficients, Standard Errors (SE), Beta's, t-values, significance, explained variance ($R^2$) and effect size ($f^2$) with 90\% confidence intervals for the chosen scaling approach on the concern for stakeholders of teams, first for an uncontrolled model and then for a model that controls for team experience with Agile and organization size. *: statistically significant at $p <.05$. **: statistically significant at $p <.01$}
\label{tab:concernForStakeholders}
\begin{tabular}{@{}m{3cm}m{1cm}m{1cm}m{1cm}m{1cm}m{1cm}m{1cm}m{2.5cm}@{}}
\toprule
\textbf{Variable} & \textbf{Estimate} & \textbf{SE} & \textbf{Beta} & \textbf{t} & \textbf{p} & \textbf{R2} & \textbf{f2 / 90\% CI} \\
\midrule
\underline{Regression model without control variables} &  &  &  &  &  & .025 & .025 [.015, .036] \\ \addlinespace
Intercept** & 3.882 & .031 & .000 & 126.132 & .000 &  &  \\ \addlinespace
Uses SAFe & .103 & .056 & .032 & 1.838 & .066 & .001 & .001 [-.001, .003] \\ \addlinespace
Uses LESS & .061 & .100 & .010 & .607 & .544 & .000 & .000 [-.001, .001] \\ \addlinespace
Uses Scrum of Scrums** & .474 & .049 & .170 & 9.615 & .000 & .022 & .023 [.013, .033] \\ \addlinespace
Custom approach & .064 & .051 & .022 & 1.243 & .214 & .000 & .000 [-.001, .002] \\ \addlinespace
Other approach** & .214 & .077 & .046 & 2.788 & .005 & .002 & .002 [-.001, .005] \\ \addlinespace
\underline{Regression model with control variables} &  &  &  &  &  & .386 & .629 [.600, .658] \\ \addlinespace
Intercept** & 1.788 & .061 & .000 & 29.290 & .000 &  &  \\ \addlinespace
Uses SAFe** & -.148 & .047 & -.046 & -3.127 & .002 & .002 & .002 [-.001, .005] \\ \addlinespace
Uses LESS & -.110 & .081 & -.018 & -1.356 & .175 & .000 & .000 [-.001, .002] \\ \addlinespace
Uses Scrum of Scrums** & .160 & .041 & .057 & 3.879 & .000 & .002 & .002 [-.001, .006] \\ \addlinespace
Custom approach* & -.083 & .042 & -.029 & -1.991 & .047 & .001 & .001 [-.001, .003] \\ \addlinespace
Other approach & .025 & .063 & .005 & .395 & .693 & .000 & .000 [.000, .000] \\ \addlinespace
Control: Team Experience with Agile** & .457 & .010 & .604 & 46.765 & .000 & .354 & .549 [.519, .578] \\ \addlinespace
Control: Size of organization** & .070 & .016 & .057 & 4.377 & .000 & .003 & .003 [-.001, .007] \\
\bottomrule
\end{tabular}
\vspace{2em}
\end{table}

\begin{table}[!ht]
\centering
\small
\caption{Coefficients, Standard Errors (SE), Beta's, t-values, significance, explained variance ($R^2$) and effect size ($f^2$) with 90\% confidence intervals for the chosen scaling approach on the continuous improvement of teams, first for an uncontrolled model and then for a model that controls for team experience with Agile and organization size. *: statistically significant at $p <.05$. **: statistically significant at $p <.01$}
\label{tab:continuousImprovement}
\begin{tabular}{@{}m{3cm}m{1cm}m{1cm}m{1cm}m{1cm}m{1cm}m{1cm}m{2.5cm}@{}}
\toprule
\textbf{Variable} & \textbf{Estimate} & \textbf{SE} & \textbf{Beta} & \textbf{t} & \textbf{p} & \textbf{R2} & \textbf{f2 / 90\% CI} \\
\midrule
\underline{Regression model without control variables} &  &  &  &  &  & .017 & .017 [.008, .026] \\ \addlinespace
Intercept** & 4.521 & .026 & .000 & 173.598 & .000 &  &  \\ \addlinespace
Uses SAFe** & .124 & .048 & .045 & 2.608 & .009 & .002 & .002 [-.001, .005] \\ \addlinespace
Uses LESS* & .197 & .085 & .038 & 2.325 & .020 & .001 & .001 [-.001, .004] \\ \addlinespace
Uses Scrum of Scrums** & .341 & .042 & .145 & 8.174 & .000 & .016 & .017 [.009, .026] \\ \addlinespace
Custom approach & .084 & .043 & .034 & 1.944 & .052 & .001 & .001 [-.001, .003] \\ \addlinespace
Other approach* & .137 & .065 & .035 & 2.099 & .036 & .001 & .001 [-.001, .003] \\ \addlinespace
\underline{Regression model with control variables} &  &  &  &  &  & .474 & .900 [.877, .928] \\ \addlinespace
Intercept** & 2.728 & .048 & .000 & 57.274 & .000 &  &  \\ \addlinespace
Uses SAFe & -.058 & .037 & -.021 & -1.578 & .115 & .000 & .000 [-.001, .002] \\ \addlinespace
Uses LESS & .042 & .063 & .008 & .670 & .503 & .000 & .000 [-.001, .001] \\ \addlinespace
Uses Scrum of Scrums & .052 & .032 & .022 & 1.630 & .103 & .000 & .000 [-.001, .002] \\ \addlinespace
Custom approach & -.047 & .033 & -.019 & -1.442 & .149 & .000 & .000 [-.001, .002] \\ \addlinespace
Other approach & -.012 & .049 & -.003 & -.250 & .802 & .000 & .000 [.000, .000] \\ \addlinespace
Control: Team Experience with Agile** & .438 & .008 & .686 & 57.478 & .000 & .458 & .846 [.817, .874] \\ \addlinespace
Control: Size of organization & -.017 & .012 & -.017 & -1.385 & .166 & .000 & .000 [-.001, .002] \\
\bottomrule
\end{tabular}
\vspace{2em}
\end{table}

\begin{table}[!ht]
\centering
\small
\caption{Coefficients, Standard Errors (SE), Beta's, t-values, significance, explained variance ($R^2$) and effect size ($f^2$) with 90\% confidence intervals for the chosen scaling approach on the autonomy of teams, first for an uncontrolled model and then for a model that controls for team experience with Agile and organization size. *: statistically significant at $p <.05$. **: statistically significant at $p <.01$}
\label{tab:autonomyOfTeams}
\begin{tabular}{@{}m{3cm}m{1cm}m{1cm}m{1cm}m{1cm}m{1cm}m{1cm}m{2.5cm}@{}}
\toprule
\textbf{Variable} & \textbf{Estimate} & \textbf{SE} & \textbf{Beta} & \textbf{t} & \textbf{p} & \textbf{R2part} & \textbf{f2 / 90\% CI} \\
\midrule
\underline{Regression model without control variables} &  &  &  &  &  & .003 & .003 [-.001, .006] \\ \addlinespace
Intercept** & 5.208 & .025 & .000 & 204.731 & .000 &  &  \\ \addlinespace
Uses SAFe & -.057 & .046 & -.022 & -1.238 & .216 & .000 & .000 [-.001, .002] \\ \addlinespace
Uses LESS & -.004 & .083 & -.001 & -.048 & .962 & .000 & .000 [.000, .000] \\ \addlinespace
Uses Scrum of Scrums* & .086 & .041 & .038 & 2.100 & .036 & .001 & .001 [-.001, .003] \\ \addlinespace
Custom approach & .044 & .042 & .018 & 1.038 & .299 & .000 & .000 [-.001, .001] \\ \addlinespace
Other approach & .069 & .064 & .018 & 1.085 & .278 & .000 & .000 [-.001, .001] \\ \addlinespace
\underline{Regression model with control variables} &  &  &  &  &  & .320 & .470 [.441, .500] \\ \addlinespace
Intercept** & 3.813 & .053 & .000 & 72.545 & .000 &  &  \\ \addlinespace
Uses SAFe** & -.185 & .041 & -.070 & -4.561 & .000 & .004 & .004 [-.001, .008] \\ \addlinespace
Uses LESS & -.119 & .070 & -.024 & -1.714 & .087 & .001 & .001 [-.001, .002] \\ \addlinespace
Uses Scrum of Scrums** & -.137 & .035 & -.060 & -3.856 & .000 & .003 & .003 [-.001, .007] \\ \addlinespace
Custom approach & -.068 & .036 & -.029 & -1.882 & .060 & .001 & .001 [-.001, .003] \\ \addlinespace
Other approach & -.054 & .054 & -.014 & -.999 & .318 & .000 & .000 [-.001, .001] \\ \addlinespace
Control: Team Experience with Agile** & .353 & .008 & .571 & 42.028 & .000 & .317 & .465 [.435, .494] \\ \addlinespace
Control: Size of organization* & -.032 & .014 & -.032 & -2.332 & .020 & .001 & .001 [-.001, .003] \\
\bottomrule
\end{tabular}
\vspace{2em}
\end{table}

\begin{table}[!ht]
\centering
\small
\caption{Coefficients, Standard Errors (SE), Beta's, t-values, significance, explained variance ($R^2$) and effect size ($f^2$) with 90\% confidence intervals for the chosen scaling approach on the management support of teams, first for an uncontrolled model and then for a model that controls for team experience with Agile and organization size. *: statistically significant at $p <.05$. **: statistically significant at $p <.01$}
\label{tab:managementSupportOfTeams}
\begin{tabular}{@{}m{3cm}m{1cm}m{1cm}m{1cm}m{1cm}m{1cm}m{1cm}m{2.5cm}@{}}
\toprule
\textbf{Variable} & \textbf{Estimate} & \textbf{SE} & \textbf{Beta} & \textbf{t} & \textbf{p} & \textbf{R2part} & \textbf{f2 / 90\% CI} \\
\midrule
\underline{Regression model without control variables} &  &  &  &  &  & .020 & .020 [.010, .030] \\ \addlinespace
Intercept** & 4.574 & .039 & .000 & 115.990 & .000 &  &  \\ \addlinespace
Uses SAFe* & .185 & .072 & .044 & 2.575 & .010 & .002 & .002 [-.001, .004] \\ \addlinespace
Uses LESS** & .496 & .129 & .062 & 3.858 & .000 & .004 & .004 [-.001, .008] \\ \addlinespace
Uses Scrum of Scrums** & .541 & .063 & .152 & 8.568 & .000 & .018 & .018 [.009, .028] \\ \addlinespace
Custom approach** & .180 & .066 & .048 & 2.742 & .006 & .002 & .002 [-.001, .005] \\ \addlinespace
Other approach** & .309 & .099 & .052 & 3.133 & .002 & .002 & .002 [-.001, .006] \\ \addlinespace
\underline{Regression model with control variables} &  &  &  &  &  & .295 & .418 [.388, .447] \\ \addlinespace
Intercept** & 2.633 & .084 & .000 & 31.430 & .000 &  &  \\ \addlinespace
Uses SAFe & .018 & .065 & .004 & .274 & .784 & .000 & .000 [.000, .000] \\ \addlinespace
Uses LESS** & .332 & .111 & .042 & 2.994 & .003 & .002 & .002 [-.001, .005] \\ \addlinespace
Uses Scrum of Scrums** & .213 & .057 & .060 & 3.774 & .000 & .003 & .003 [-.001, .007] \\ \addlinespace
Custom approach & .044 & .057 & .012 & .771 & .440 & .000 & .000 [-.001, .001] \\ \addlinespace
Other approach & .166 & .086 & .028 & 1.931 & .054 & .001 & .001 [-.001, .003] \\ \addlinespace
Control: Team Experience with Agile** & .517 & .013 & .533 & 38.540 & .000 & .276 & .381 [.352, .411] \\ \addlinespace
Control: Size of organization** & -.089 & .022 & -.057 & -4.048 & .000 & .003 & .003 [-.001, .007] \\
\bottomrule
\end{tabular}
\vspace{2em}
\end{table}

\begin{table}[!ht]
\centering
\small
\caption{Coefficients, Standard Errors (SE), Beta's, t-values, significance, explained variance ($R^2$) and effect size ($f^2$) with 90\% confidence intervals for the chosen scaling approach on team effectiveness, first for an uncontrolled model and then for a model that controls for team experience with Agile and organization size. *: statistically significant at $p <.05$. **: statistically significant at $p <.01$}
\label{tab:teamEffectiveness}
\begin{tabular}{@{}m{3cm}m{1cm}m{1cm}m{1cm}m{1cm}m{1cm}m{1cm}m{2.5cm}@{}}
\toprule
\textbf{Variable} & \textbf{Estimate} & \textbf{SE} & \textbf{Beta} & \textbf{t} & \textbf{p} & \textbf{R2part} & \textbf{f2 / 90\% CI} \\
\midrule
\underline{Regression model without control variables} &  &  &  &  &  & .012 & .012 [.004, .019] \\ \addlinespace
Intercept** & 5.031 & .029 &  & 176.450 & .000 &  &  \\ \addlinespace
Uses SAFe** & .138 & .052 & .046 & 2.656 & .008 & .002 & .002 [-.001, .005] \\ \addlinespace
Uses LESS & .037 & .093 & .006 & .399 & .690 & .000 & .000 [.000, .000] \\ \addlinespace
Uses Scrum of Scrums** & .309 & .046 & .121 & 6.778 & .000 & .011 & .011 [.004, .019] \\ \addlinespace
Custom approach* & .116 & .047 & .043 & 2.456 & .014 & .001 & .001 [-.001, .004] \\ \addlinespace
Other approach* & .156 & .071 & .036 & 2.193 & .028 & .001 & .001 [-.001, .004] \\ \addlinespace
\underline{Regression model with control variables} &  &  &  &  &  & .384 & .624 [.595, .653] \\ \addlinespace
Intercept** & 3.205 & .056 &  & 57.001 & .000 &  &  \\ \addlinespace
Uses SAFe & -.060 & .043 & -.020 & -1.369 & .171 & .000 & .000 [-.001, .002] \\ \addlinespace
Uses LESS & -.093 & .075 & -.016 & -1.243 & .214 & .000 & .000 [-.001, .001] \\ \addlinespace
Uses Scrum of Scrums & .025 & .038 & .010 & .659 & .510 & .000 & .000 [-.001, .001] \\ \addlinespace
Custom approach & -.019 & .039 & -.007 & -.505 & .614 & .000 & .000 [.000, .001] \\ \addlinespace
Other approach & -.012 & .058 & -.003 & -.212 & .832 & .000 & .000 [.000, .000] \\ \addlinespace
Control: Team Experience with Agile** & .430 & .009 & .618 & 47.802 & .000 & .371 & .591 [.562, .620] \\ \addlinespace
Control: Size of organization & .011 & .015 & .009 & .719 & .472 & .000 & .000 [-.001, .001] \\
\bottomrule
\end{tabular}
\vspace{2em}
\end{table}

\begin{table}[!ht]
\centering
\small
\caption{Coefficients, Standard Errors (SE), Beta's, t-values, significance, explained variance ($R^2$) and effect size ($f^2$) with 90\% confidence intervals for the chosen scaling approach on stakeholder satisfaction with value, first for an uncontrolled model and then for a model that controls for team experience with Agile and organization size. *: statistically significant at $p <.05$. **: statistically significant at $p <.01$}
\label{tab:stakeholderSatisfactionWithValue}
\begin{tabular}{@{}m{3cm}m{1cm}m{1cm}m{1cm}m{1cm}m{1cm}m{1cm}m{2.5cm}@{}}
\toprule
\textbf{Variable} & \textbf{Estimate} & \textbf{SE} & \textbf{Beta} & \textbf{t} & \textbf{p} & \textbf{R2part} & \textbf{f2 / 90\% CI} \\
\midrule
\underline{Regression model without control variables} &  &  &  &  &  & .028 & .028 [.003, .059] \\ \addlinespace
Intercept** & 5.327 & .061 &  & 87.173 & .000 &  &  \\ \addlinespace
Uses SAFe & .170 & .139 & .055 & 1.220 & .223 & .003 & .003 [-.006, .013] \\ \addlinespace
Uses LESS & .433 & .284 & .067 & 1.526 & .128 & .004 & .004 [-.006, .017] \\ \addlinespace
Uses Scrum of Scrums* & .260 & .113 & .107 & 2.296 & .022 & .010 & .010 [-.006, .029] \\ \addlinespace
Custom approach & .185 & .112 & .077 & 1.656 & .098 & .005 & .005 [-.006, .019] \\ \addlinespace
Other approach** & .480 & .145 & .150 & 3.317 & .001 & .020 & .021 [-.002, .048] \\ \addlinespace
\underline{Regression model with control variables} &  &  &  &  &  & .091 & .101 [.057, .153] \\ \addlinespace
Intercept** & 4.136 & .215 &  & 19.205 & .000 &  &  \\ \addlinespace
Uses SAFe & .032 & .141 & .011 & .230 & .818 & .000 & .000 [-.001, .002] \\ \addlinespace
Uses LESS & .479 & .274 & .074 & 1.749 & .081 & .005 & .005 [-.006, .019] \\ \addlinespace
Uses Scrum of Scrums & .120 & .111 & .049 & 1.083 & .279 & .002 & .002 [-.005, .011] \\ \addlinespace
Custom approach & .123 & .108 & .051 & 1.132 & .258 & .002 & .002 [-.005, .011] \\ \addlinespace
Other approach** & .424 & .140 & .133 & 3.025 & .003 & .016 & .016 [-.004, .040] \\ \addlinespace
Control: Team Experience with Agile** & .253 & .038 & .284 & 6.632 & .000 & .076 & .082 [.041, .130] \\ \addlinespace
Control: Size of organization & -.019 & .044 & -.019 & -.431 & .667 & .000 & .000 [-.003, .004] \\
\bottomrule
\end{tabular}
\vspace{2em}
\end{table}

\begin{table}[!ht]
\centering
\small
\caption{Coefficients, Standard Errors (SE), Beta's, t-values, significance, explained variance ($R^2$) and effect size ($f^2$) with 90\% confidence intervals for the chosen scaling approach on stakeholder satisfaction with responsiveness, first for an uncontrolled model and then for a model that controls for team experience with Agile and organization size. *: statistically significant at $p <.05$. **: statistically significant at $p <.01$}
\label{tab:stakeholderSatisfactionWithResponsiveness}
\begin{tabular}{@{}m{3cm}m{1cm}m{1cm}m{1cm}m{1cm}m{1cm}m{1cm}m{2.5cm}@{}}
\toprule
\textbf{Variable} & \textbf{Estimate} & \textbf{SE} & \textbf{Beta} & \textbf{t} & \textbf{p} & \textbf{R2part} & \textbf{f2 / 90\% CI} \\
\midrule
\multicolumn{3}{l}{\underline{Regression model without control variables}} &  &  &  & N/A & N/A \\ \addlinespace
Model not significant &  &  &  &  &  &  &  \\ \addlinespace
\underline{Regression model with control variables} &  &  &  &  &  & .064 & .068 [.030, .113] \\ \addlinespace
Intercept** & 4.318 & .217 & .000 & 19.859 & .000 &  &  \\ \addlinespace
Uses SAFe & .027 & .142 & .009 & .193 & .847 & .000 & .000 [-.001, .002] \\ \addlinespace
Uses LESS & .220 & .276 & .034 & .795 & .427 & .001 & .001 [-.004, .008] \\ \addlinespace
Uses Scrum of Scrums & .201 & .112 & .084 & 1.795 & .073 & .006 & .006 [-.006, .020] \\ \addlinespace
Custom approach & -.031 & .109 & -.013 & -.286 & .775 & .000 & .000 [-.002, .002] \\ \addlinespace
Other approach & .190 & .142 & .060 & 1.344 & .180 & .003 & .003 [-.006, .014] \\ \addlinespace
Control: Team Experience with Agile** & .187 & .038 & .213 & 4.868 & .000 & .042 & .044 [.013, .082] \\ \addlinespace
Control: Size of organization & .049 & .044 & .050 & 1.117 & .264 & .002 & .002 [-.005, .011] \\
\bottomrule
\end{tabular}
\vspace{2em}
\end{table}

\begin{table}[!ht]
\centering
\small
\caption{Coefficients, Standard Errors (SE), Beta's, t-values, significance, explained variance ($R^2$) and effect size ($f^2$) with 90\% confidence intervals for the chosen scaling approach on stakeholder satisfaction with release frequency, first for an uncontrolled model and then for a model that controls for team experience with Agile and organization size. *: statistically significant at $p <.05$. **: statistically significant at $p <.01$}
\label{tab:stakeholderSatisfactionWithReleaseFrequency}
\begin{tabular}{@{}m{3cm}m{1cm}m{1cm}m{1cm}m{1cm}m{1cm}m{1cm}m{2.5cm}@{}}
\toprule
\textbf{Variable} & \textbf{Estimate} & \textbf{SE} & \textbf{Beta} & \textbf{t} & \textbf{p} & \textbf{R2part} & \textbf{f2 / 90\% CI} \\
\midrule
\multicolumn{3}{l}{\underline{Regression model without control variables}} &  &  &  & N/A & N/A \\ \addlinespace
Model not significant &  &  &  &  &  &  &  \\ \addlinespace
\underline{Regression model with control variables} &  &  &  &  &  & .081 & .088 [.046, .130] \\ \addlinespace
Intercept** & 3.814 & .258 & .000 & 14.774 & .000 &  &  \\ \addlinespace
Uses SAFe & -.074 & .168 & -.020 & -.438 & .661 & .000 & .000 [-.003, .003] \\ \addlinespace
Uses LESS & .488 & .328 & .063 & 1.486 & .138 & .004 & .004 [-.006, .014] \\ \addlinespace
Uses Scrum of Scrums & .024 & .133 & .008 & .178 & .859 & .000 & .000 [-.001, .001] \\ \addlinespace
Custom approach & .051 & .130 & .018 & .395 & .693 & .000 & .000 [-.002, .003] \\ \addlinespace
Other approach & .225 & .168 & .060 & 1.340 & .181 & .003 & .003 [-.006, .012] \\ \addlinespace
Control: Team Experience with Agile** & .290 & .046 & .275 & 6.347 & .000 & .071 & .077 [.037, .116] \\ \addlinespace
Control: Size of organization & -.047 & .052 & -.039 & -.888 & .375 & .001 & .001 [-.005, .007] \\
\bottomrule
\end{tabular}
\vspace{2em}
\end{table}

\end{document}